\documentclass[twocolumn,aps,prl,shownopacs,superscriptaddress]{revtex4}
\usepackage{graphicx}
\usepackage{amssymb}
\usepackage{amsmath}
\usepackage{amsthm}
\usepackage{newtxtext,newtxmath}
\usepackage{dsfont}
\usepackage{wasysym}
\usepackage{color}
\usepackage{array}
\usepackage{multirow}
\usepackage{makecell}
\usepackage{xcolor}
\usepackage[colorlinks=true,linkcolor=blue,urlcolor=blue,citecolor=blue,pdfauthor={ },pdftitle={},pdfsubject={ },pdfkeywords={ }]{hyperref}
\graphicspath{{Figures/}}

\begin{document}
	
	\title{Variational Quantum Metrology with Loschmidt Echo}
	
	\date{\today}
\author{Ran Liu}
\thanks{These two authors contributed equally}
\affiliation{CAS Key Laboratory of Microscale Magnetic Resonance and School of Physical Sciences, University of Science and Technology of China, Hefei 230026, China}
\affiliation{CAS Center for Excellence in Quantum Information and Quantum Physics, University of Science and Technology of China, Hefei 230026, China}
\affiliation{Institute of Quantum Precision Measurement, State Key Laboratory of Radio Frequency Heterogeneous Integration, College of Physics and Optoelectronic Engineering, Shenzhen University, Shenzhen 518060, China}
\author{Ze Wu}
\thanks{These two authors contributed equally}
\affiliation{CAS Key Laboratory of Microscale Magnetic Resonance and School of Physical Sciences, University of Science and Technology of China, Hefei 230026, China}
\affiliation{CAS Center for Excellence in Quantum Information and Quantum Physics, University of Science and Technology of China, Hefei 230026, China}
\author{Xiaodong Yang}
\affiliation{Institute of Quantum Precision Measurement, State Key Laboratory of Radio Frequency Heterogeneous Integration, College of Physics and Optoelectronic Engineering, Shenzhen University, Shenzhen 518060, China}
\affiliation{Quantum Science Center of Guangdong-Hong Kong-Macao Greater Bay Area (Guangdong), Shenzhen 518045, China}
\author{Yuchen Li}
\affiliation{CAS Key Laboratory of Microscale Magnetic Resonance and School of Physical Sciences, University of Science and Technology of China, Hefei 230026, China}
\affiliation{CAS Center for Excellence in Quantum Information and Quantum Physics, University of Science and Technology of China, Hefei 230026, China}
\author{Hui Zhou}
\affiliation{School of Physics, Hefei University of Technology, Hefei, Anhui 230009, China}
\author{Zhaokai Li}
	\affiliation{CAS Key Laboratory of Microscale Magnetic Resonance and School of Physical Sciences, University of Science and Technology of China, Hefei 230026, China}
	\affiliation{CAS Center for Excellence in Quantum Information and Quantum Physics, University of Science and Technology of China, Hefei 230026, China}
	\affiliation{Hefei National Laboratory, University of Science and Technology of China, Hefei 230088, China}
\author{Yuquan Chen}
\affiliation{CAS Key Laboratory of Microscale Magnetic Resonance and School of Physical Sciences, University of Science and Technology of China, Hefei 230026, China}
\affiliation{CAS Center for Excellence in Quantum Information and Quantum Physics, University of Science and Technology of China, Hefei 230026, China}

\author{Haidong Yuan}
\email{hdyuan@mae.cuhk.edu.hk}
\affiliation{Department of Mechanical and Automation Engineering, The Chinese University of Hong Kong, Shatin, Hong Kong SAR, China}
\author{Xinhua Peng}
\email{xhpeng@ustc.edu.cn}
\affiliation{CAS Key Laboratory of Microscale Magnetic Resonance and School of Physical Sciences, University of Science and Technology of China, Hefei 230026, China}
\affiliation{CAS Center for Excellence in Quantum Information and Quantum Physics, University of Science and Technology of China, Hefei 230026, China}
\affiliation{Hefei National Laboratory, University of Science and Technology of China, Hefei 230088, China}
	
\begin{abstract}
	By leveraging quantum effects, such as superposition and entanglement, quantum metrology promises higher precision than the classical strategies. It is, however, a challenging task to achieve the higher precision on practical systems. This is mainly due to the difficulties in engineering non-classical states and performing nontrivial measurements on the system, especially when the number of particles is large. Here we propose a variational scheme with Loschmidt echo for quantum metrology. By utilizing hardware-efficient Ans\"{a}tze in the design of variational quantum circuits, the quantum Fisher information (QFI) of the probe state can be extracted from the experimentally measured Loschmidt echo in a scalable manner. This QFI is then used to guide the online optimization of the preparation of the probe state. We experimentally implement the scheme on an ensemble of 10-spin quantum processors and achieve a 12.4 dB enhancement of the measurement precision over the uncorrelated states, which is close to the theoretical limit.  The scheme can also be employed on various other noisy intermediate-scale quantum devices, which provides a promising protocol to demonstrate quantum advantages.     

\end{abstract}
\maketitle
	
	To sense more accurately has always been one of the main drives for scientific advances and technological innovations. Quantum metrology \cite{Qesti1,Qesti2,Qesti3}, which utilizes quantum correlations to achieve higher sensitivities, has gained much attention recently. In ideal scenarios, quantum metrology can achieve a precision at the Heisenberg limit, which scales as $1/N$ with $N$ as the number of particles \cite{advanceQM,QM,HL1,RewSensing}. As a contrast, the precision of the classical strategies is bounded by the standard quantum limit (SQL), which scales as $1/\sqrt{N}$. To achieve the higher precisions in quantum metrology, nontrivial entangled probe states, however, need to be prepared. This poses a practically challenging task when the number of particles increases. In practice, there are two main difficulties in achieving the highest precision. First, it is difficult to identify the optimal probe state when the number of particles increases. Due to the "curse of dimensionality", the classical optimization that is required to identify the optimal probe state soon becomes intractable \cite{Yuan_opt,Qiushi2023,yuan2017quantum,Mao2022}. Second, it is a challenging task to prepare the identified optimal probe state on practical systems due to the device-specific constraints, such as decoherences, imperfect controls and readout errors\cite{Noise1,Noise2,Noise4}.
	
	Variational quantum metrology (VQM) provides a promising route to circumvent these problems. In VQM the identification of the optimal probe state is carried out with a hybrid quantum-classical scheme. A variational quantum circuit is used to prepare the probe state and the circuit is optimized externally by a classical computer \cite{VQM1,VQM2,VQM3,Yang}. This hybrid scheme inherits the advantages of the variational quantum algorithm which not only reduces the complexity of the classical simulation, but can also easily incorporate the device-specific constraints into the design of the variational quantum circuit (VQC). The optimization of the circuit, however, can still be very challenging for quantum metrology. This is because the quantum Fisher information (QFI), which is often taken as the figure of merit in quantum metrology, is difficult to evaluate. 
 The general brute-force approaches to extract QFI, such as quantum state tomography, demand an exponentially growing number of measurements \cite{nielsen2002}. Although some effective surrogates of QFI have been proposed previously, such as those based on additional physical qubits or experimental measurements \cite{FQ_tim1,FQ_tim2,Yang,FQ_num1,FQ_num2,FQ_num3}, they may still require a considerable number of experimental measurements or extra physical qubits. This can go beyond the current experimental capabilities.

	In this Letter, we propose a variational optimization scheme for quantum metrology that uses Loschmidt echo (LE) to efficiently extract the QFI. The signal of the LE can then be directly used to optimize the VQC which prepares the optimal probe state in quantum metrology. We demonstrate the power of the scheme by identifying and preparing a 10-spin optimal probe state in nuclear magnetic resonance (NMR) for the estimation of an unknown phase, where the system is in mixed states at room temperature. We experimentally implement the scheme and demonstrate that the achieved precision is close to the fundamental bound in quantum metrology---the quantum Cram\'er-Rao bound (QCRB). This opens a promising avenue for the implementation of quantum enhanced parameter estimation on practical quantum devices due to its efficiency, robustness against experimental imperfections and easy implementation.   
	\begin{figure}
		\begin{center}
			\includegraphics[scale=1]{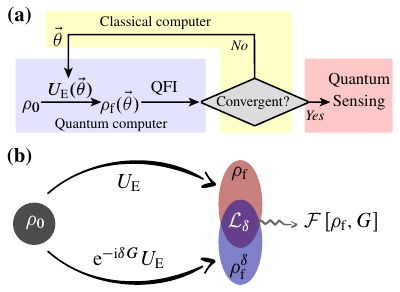} 
			\caption{(a).\ Workflow for quantum probe engineering via quantum variational optimization. By taking the QFI as the figure of merit for the optimization of VQC, the probe state is steered to the optimal state for high precision phase estimation under practical dynamics. (b).\ Schematic diagram of measuring LE. When the unperturbed evolution is specified as the engineering operation, i.e., $U\to U_\text{E}(\vec{\theta})$, and the perturbation in the perturbed evolution is specified as a small quench under encoding dynamics, i.e., $U_\delta\to e^{-i\delta G}U_\text{E}(\vec{\theta})$, respectively, the QFI of the engineered probe $\rho_\text{f}$ can then be extracted from LE.} \label{diagram}
		\end{center}
	\end{figure}

\section{RESULTS}
\subsection{Scheme}
We consider the iconic task of estimating the parameter $\alpha$ in the operator $U_\alpha=e^{-i\alpha G}$ with $G$ as the generator. The ultimate precision can be quantified by the QCRB, \cite{Qesti1,Qesti2,Qesti3,QCRB1,QCRB2} as
	\begin{eqnarray}\label{CRb}
		\Delta\alpha\ge\frac{1}{\sqrt{\nu\mathcal{F}}},
	\end{eqnarray}
	where  $\Delta\alpha$ is the standard deviation of an unbiased estimator $\hat{\alpha}$, $\nu$ is the number of repetitive measurements, $\mathcal{F}$ is the QFI.
	Our target here is to engineer a probe state with the maximal QFI, which leads to the smallest standard deviation. Here, the probe state is prepared by a VQC, which generates a unitary operation, $U_\text{E}(\vec{\theta})$, acting on a natural initial state of the physical system with $\vec{\theta}$ being the tunable parameters of the circuit. By taking the QFI as the figure of merit we then optimize $\vec{\theta}$ to steer the probe state towards the optimal or nearly optimal state. This state is subsequently used for high-precision phase estimation. The schematic for the workflow of the quantum probe engineering via VQM is illustrated in Fig.\ \ref{diagram}(a).

An essential part of the variational optimization is to efficiently evaluate the figure of merit that determines how the parameters should be tuned. However, the standard methods of evaluating the QFI, such as state tomography, are extremely demanding in experiments. Here we develop an experimental protocol that uses Loschmidt echo to evaluate the QFI. 

\begin{figure}
		\begin{center}
			\includegraphics[scale=0.75]{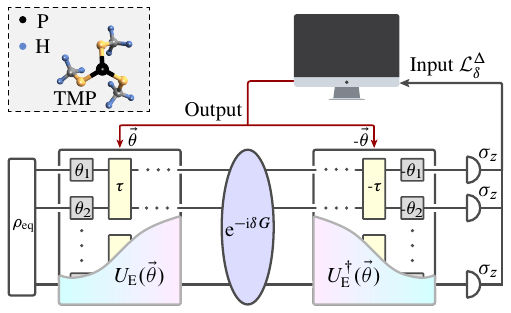} 
			\caption{The experimental procedures for  variationally optimizing the metrologically useful mixed state via LE. The 10-spin quantum probe is realized by a $^{31}\text{P}$ nuclear spin and nine equivalent $^1\text{H}$ spins in the TMP molecule and initialized as an equilibrium state $\rho_\text{eq}$. $\rho_\text{eq}$  evolves under the symmetrical variational quantum circuit $V_\delta(\vec{\theta})\equiv U_\text{E}^\dagger(\vec{\theta}) e^{-i\delta G}U_\text{E}(\vec{\theta})$ and the polarization of each spin along $z$-axis then is measured to obtain the LE $\mathcal L_\delta$. The QFI of quantum probe, i.e., $\mathcal F\left[\rho_\text{f}(\vec{\theta}),G\right]$, can be extracted from LE and feedback to the classical computer, which is employed to iteratively update the parameters $\vec{\theta}$ to maximize the QFI.}\label{exppro}
		\end{center}
\end{figure}

For pure state the Loschmidt echo is given by $\mathcal L_\delta=\left|\left\langle\Psi_{0}\left|U^\dagger U_\delta\right| \Psi_{0}\right\rangle\right|^{2}$, which is the overlap between the states obtained from the forward unperturbed evolution $\left(U\right)$ and the forward perturbed evolution $\left(U_\delta\right)$. The Loschmidt echo corresponds to a susceptibility to the perturbation \cite{LE_def,LE_def2}. As shown in Fig.\ \ref{diagram}(b), the Loschmidt echo can be used to extract the QFI when we substitute $U$ and $U_\delta$ with $U_\text{E}(\vec{\theta})$ and $e^{-i\delta G}U_\text{E}(\vec{\theta})$ respectively. In this case, the Fisher information can be evaluated from the Loschmidt echo as\cite{QFILE}
	\begin{eqnarray}\label{eq:LEQFI}
		\mathcal F\left[U_\text{E}(\vec{\theta})|\Psi_0\rangle\right]=\lim_{\delta\to0}4\frac{1-\mathcal{L}_{\delta}}{\delta^2}.
	\end{eqnarray}
We generalize this connection to the initially mixed quantum system, in which the considered process for state preparation is still unitary. The Loschmidt echo then becomes
	\begin{eqnarray}\label{eqLE}
		\begin{aligned}
		\mathcal L_\delta\equiv&\text{Tr}\left[\rho_\text{f}\rho_\text{f}^\delta\right]\\
		\approx&\Gamma(\rho_\text{f})-\frac{\delta^2}{4}\times\left[2\sum_{i,j=1}^d(\lambda_i-\lambda_j)^2|\langle\psi_i|G|\psi_j\rangle|^2\right].
		\end{aligned}
	\end{eqnarray}
	Here $\rho_\text{f}=U_\text{E}(\vec{\theta})\rho_0U_\text{E}^\dagger(\vec{\theta})=\sum_{i=1}^d\lambda_i$$|\psi_i\rangle\langle\psi_i|$,$\rho_\text{f}^\delta=e^{-i\delta G}U_\text{E}(\vec{\theta})\rho_0U_\text{E}^\dagger(\vec{\theta}) e^{i\delta G}$, $d$ is the dimension of the Hilbert space, $\Gamma(\rho_\text{f})=\sum_{i=1}^d\lambda_i^2$ is the purity of the state, which in our case, can be treated as a constant since it does not change under unitary evolution. LE is connected to the QFI of  mixed states as (see Methods for detailed deviations)
	\begin{eqnarray}\label{ineqFq}
		\mathcal F\left[\rho_\text{f},H\right]\ge\lim_{\delta\to0}4\frac{\Gamma(\rho_\text{f})-\mathcal L_\delta}{\delta^2}.
	\end{eqnarray}
    Though only a lower bound of QFI can be extracted from this inequality, this bound is directly related to sub-QFI, which shares the same global extrema with QFI \cite{cerezo2021} and can thus be employed to the variational optimization of probe state. For highly mixed states where the eigenvalues are almost degenerate, i.e., $\lambda_i\approx\frac{1}{d}$ for $1\le i\le d$, the bound can also be saturated with
	\begin{eqnarray}\label{ineqFq2}
		\mathcal F\left[\rho_\text{f},H\right]\approx\lim_{\delta\to0}2d\frac{\Gamma(\rho_\text{f})-\mathcal L_\delta}{\delta^2}.
	\end{eqnarray}
    This is exactly the case in NMR as the initial state of the NMR system is a thermal state with the Boltzmann distribution, which at the room-temperature is close to the completely mixed state \cite{spindyn}. Since $\rho_\text{f}=U_\text{E}(\vec{\theta})\rho_0U_\text{E}^\dagger(\vec{\theta})$ has the same eigenvalue as $\rho_0$, $\rho_\text{f}$ is thus also almost degenerate.

	For a better understanding of the experimental extraction of LE, we would like to rewrite Eq.\ \eqref{eqLE} as 
	\begin{eqnarray}\label{eqLEexp}
	\mathcal L_\delta\equiv\text{Tr}\left[V_\delta(\vec{\theta})\rho_0V_\delta^\dagger(\vec{\theta})\rho_0\right]
	\end{eqnarray}
	with $V_\delta(\vec{\theta})\equiv U_\text{E}^\dagger(\vec{\theta}) e^{-i\delta G}U_\text{E}(\vec{\theta})$. The LE can thus be obtained by first using the variational quantum circuit to generate $U_\text{E}(\vec{\theta})$, then apply a perturbation evolution $e^{-i\delta H}$, followed by a backward evolution $U^\dagger_\text{E}(\vec{\theta})$ and a projection onto the initial state. 
	
We note that the initial states of practical quantum systems are typically classical product states, making the LE efficiently extractable from linearly increasing local measurements with the system size and experimentally favorable. Moreover, the VQCs can be designed with hardware-efficient Ans\"{a}tze \cite{kandala2017}, which not only enhance their feasibility across diverse quantum systems by accommodating the constraints of current quantum hardware but also ensure that the backward evolution  $U^\dagger_\text{E}(\vec{\theta})$ can be implemented in a scalable manner \cite{supp}.
	
\begin{figure}
		\begin{center}
			\includegraphics[scale=0.8]{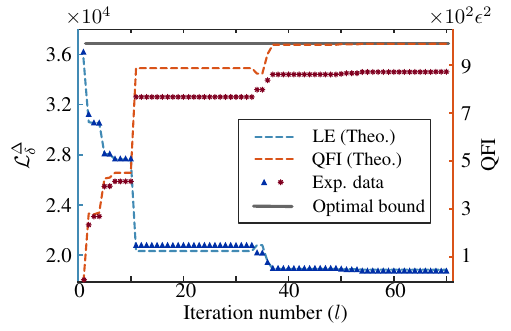} 
			\caption{Experimental results of variational quantum optimization. The blue triangles are the measured $\mathcal L_\delta^{\Delta}$ in the experiment. Error bars are absent due to the fluctuations being smaller than the size of the points \cite{supp}. The blue dashed line is the theoretical LE obtained from numerical calculation. The estimated QFI according to Eq.~\eqref{ineqFq2} is depicted with the red stars, while the red dashed line is the theoretical QFI. The theoretical maximum of QFI given by Ref. \cite{MaxFq} is $\mathcal F_\text{max}=989\epsilon^2$ and plotted with the black solid line. The finally engineered probe is close to the optimal one even in the presence of experimental imperfections. To compensate for the signal decay caused by relaxation, the experimental results of LE and QFI have been calibrated.} \label{fig2}
		\end{center}
	\end{figure}

\subsection{Experimental variational optimization of 10-spin mixed quantum probe state}

	We experimentally demonstrate the scheme on a Bruker Avance III 400 MHz NMR spectrometer at room temperature. The sample is trimethylphosphite (TMP) dissolved in $d_6$ acetone. The TMP molecule, which consists of a central $^{31}\text{P}$ nuclear spin and nine equivalent $^{1}\text{H}$ nuclear spins as shown in Fig.~\ref{exppro}, is employed as the 10-spin quantum probe. In the liquid state, the interaction between $^{1}\text{H}$ spins is negligible due to the magnetic equivalence. The natural Hamiltonian of the system in doubly rotating frame is $H_{\text{NMR}}=\frac{\pi}{2}J_{\text{PH}}\sigma_z^1\otimes\sum_{j=2}^{10}\sigma_z^j$ with $J_\text{PH}=10.5$ Hz. Here a $^{31}\text{P}$ nuclear spin and nine $^1\text{H}$ nuclear spins are denoted by Arabic numerals, respectively.

    Fig.\ \ref{exppro} shows the experimental procedures for engineering the mixed probe state via the hybrid quantum-classical scheme with quantum variational optimization. Here the extraction of LE is performed on the quantum system, while the updating of the parameters is determined on the classical computer. The quantum part contains three major stages as described below.
	
(i) The system is initially at the uncorrelated equilibrium state of the room temperature, $\rho_\text{eq}=(\mathds{1}+\epsilon\rho_\text{eq}^\Delta)/2^{10}$ where $\rho_\text{eq}^\Delta=\sum_{j=1}^{10}\gamma_{j}\sigma_z^{j}/2$, $\mathds{1}$ is the $2^{10}\times2^{10}$ unit operator, $\epsilon$ is the thermal polarization ($\sim10^{-5}$) and $\gamma_{j}$ is the relative gyromagnetic ratio of the corresponding nuclear with $\gamma_1=0.8,\gamma_{2,3,...10}=2.0$.

(ii) Evolve the system under $V_\delta(\vec{\theta})\equiv U_\text{E}^\dagger(\vec{\theta}) e^{-i\delta G}U_\text{E}(\vec{\theta})$ according to Eq.~\eqref{eqLEexp}. In our experiment, $U_\text{E}(\vec{\theta})$ is realized by a 3-layer VQC consisting of single-spin rotations, i.e., $e^{-i\theta_k\sigma_{x,y}/2}$ with $\vec\theta\equiv(\theta_1,\theta_2,...\theta_k,...)$, and the free evolution under the Hamiltonian, $H_\text{NMR}$ for a duration $\tau$. The interactions in $H_\text{NMR}$ facilitate the generation of nonclassical correlations in the probe state, thereby enabling the potential to achieve precision beyond the SQL. Details of VQC can be found in the Supplemental Materials \cite{supp}. After the preparation of the optimal probe state, the dynamics that encodes the parameter $e^{-i\delta G}$ is then applied. Without loss of generality, we consider the encoding dynamics as a field along $z$-axis and the corresponding Hamiltonian is  $G=\sum_{k=1}^{10}\sigma_z^j/2$. Theoretically, the correspondence between the Loschmidt echo and the QFI is best when $\delta\to0$, as indicated in Eq. \eqref{ineqFq2}. However, the experiment signal of the Loschmidt echo is least sensitive to the change of the parameter when $\delta=0$ since $\mathcal L_\delta=\Gamma(\rho_f)$ reaches the maximal where the derivative is zero. So there exists a trade-off. With the aid of numerical simulation, we find that $\delta=0.2$ is optimal for our experiment \cite{supp}. Finally, the reverse evolution $U^\dagger_\text{E}(\vec{\theta})$ is performed. This can be implemented by applying the reverse evolution of each operation in the PQC in reverse order. Specifically, the reverse of single-spin rotations can be implemented by changing the phase of each pulse, and the reverse of the free evolution under $H_\text{NMR}$ can be implemented by applying $\pi$ pulses along the $x$-direction to the $^{31}$P spin at both the beginning and end of the evolution.

 (iii) Project the evolved state onto the initial state $\rho_0$. Substituting the specific form of $\rho_0$ into Eq.~\eqref{eqLEexp}, we have 
\begin{equation}\label{eqLEexp2}
 	\mathcal L_\delta=\frac{1}{2^N}+\frac{\epsilon}{2^N}\sum_{j=1}^{10}\gamma_j\text{Tr}\left[V_\delta(\vec{\theta})\rho_0V_\delta^\dagger(\vec{\theta})\sigma_z^j\right].
\end{equation}
This means LE can be extracted from the local measurement of the evolved state $V_\delta(\vec{\theta})\rho_0V_\delta^\dagger(\vec{\theta})$, i.e., the polarization of each spin along $z$-axis. Hence the measurement overhead increases linearly with the system size. The identity $\mathds{1}$ in $\rho_0$ does not change under the unitary evolution $V_\delta(\vec{\theta})$ and also does not contribute to the experimental signal since the observables in NMR are traceless. The Loschmidt echo in Eq.~\eqref{eqLEexp2} then becomes
	\begin{equation}
	\mathcal L_\delta=\frac{1}{2^N}+\frac{\epsilon^2}{2^{2N}}\mathcal L_\delta^\Delta
	\end{equation}
 with $\mathcal L_\delta^\Delta\equiv\sum_{j=1}^{10}\gamma_j\text{Tr}\left(V_\delta(\vec{\theta})\rho_\text{eq}^\Delta V^\dagger_\delta(\vec{\theta})\sigma_z^j\right)$, and $\text{Tr}\left(V_\delta(\vec{\theta})\rho_\text{eq}^\Delta V^\dagger_\delta(\vec{\theta})\sigma_z^j\right)$ is directly obtained from the experimental measurements on different nuclear spins.

To reduce errors in measuring LE, we employ several techniques in our experiment. We use single-spin rotations with the BB1 composited sequence \cite{NMR_BB1} to address pulse shape imperfections. To enhance the signal-to-noise ratio, protons are decoupled during the measurement of the $^{31}\text{P}$ nucleus signals. The total evolution duration is 19 ms, whereas the decoherence time is 44 ms. The signal decay due to decoherence is therefore non-negligible. We compensate for this decay by calibrating the signal $\mathcal L_\delta^{\Delta}$ using $\mathcal L_0^\Delta\equiv\sum_{j=1}^{10}\gamma_j\text{Tr}\left(V_0(\vec{\theta})\rho_\text{eq}^\Delta V^\dagger_0(\vec{\theta})\sigma_z^j\right)$, which has a known theoretical value and a similar level of decay as $\mathcal L_\delta^{\Delta}$ \cite{supp}. 
	\begin{figure}
		\begin{center}
			\includegraphics[scale=0.8]{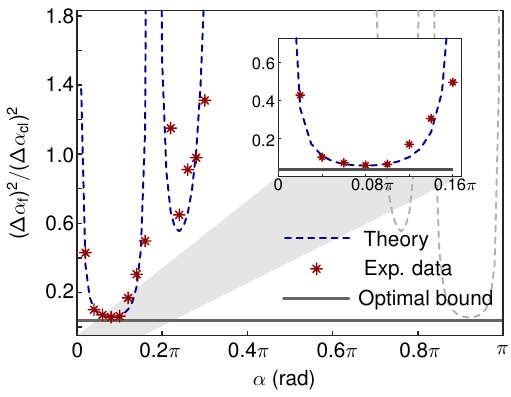} 
			\caption{The percision ratio of experimentally engineered state $\rho_\text{f}$ to the classical uncorrelated state $\rho_\text{cl}$. These experimental results have been calibrated to compensate for the signal decay caused by relaxation. The red stars are obtained from experimental measurements of $\rho_\text{f}$, which outperforms the precision limit of $\rho_\text{cl}$ with a factor of 12.4 dB (a factor of 10.7 dB without signal compensation), and the blue dashed line is the theoretical result. The gray solid line is the optimal precision given by QFI and bounds the precision of experimental measurements. It is more clear in the inset that though the time-reversal-based readout protocol is suboptimal, a precision close to the optimal QCRB can still be realized with current engineered probe.} \label{fig3}
		\end{center}
	\end{figure}
	
	With the extracted LE from our quantum processor, we proceed to train the parameters in the PQC using a classical optimizer. Specifically, we adopt the Nelder-Mead (NM) algorithm \cite{NMalgo} due to its enhanced robustness against noise and ability to explore neighboring valleys to identify better local optima. These characteristics make the NM algorithm particularly well-suited for our experimental implementation. We have also made modifications to the algorithm to further improve its efficiency \cite{supp}.

	The experimental results are illustrated in Fig.~\ref{fig2}, where the blue triangles represent the measured $\mathcal L_\delta^{\Delta}$ obtained in the experiment. It is observed that the signal initially drops rapidly and tends to stabilize with increasing iteration number $l$, capped at a maximum of 70. The blue dashed line shows the theoretical LE signal, serving as a benchmark for experimental accuracy. The relative error between the theoretical and experimental results is $1.35\%$, primarily attributed to relaxation effects. A detailed analysis of experimental error is put in the Supplemental Materials \cite{supp}.  The QFI extracted from the experimental data using Eq.~\eqref{ineqFq2} is denoted by the red stars, while the theoretical QFI is represented by the red dashed line. The discrepancy between them arises from the experimental errors and the neglected higher-order terms in the correspondence between the Loschmidt echo and QFI at finite $\delta$. Additionally, the optimal QFI predicted by Ref. \cite{MaxFq} is plotted with a solid black line. While the theoretical QFI does not always increase as the experimental one, for instance at $l=34$ due to the experimental error, it still converges to a significantly enhanced QFI close to the maximum. This result validates the feasibility of our scheme in the presence of experimental imperfections.

\subsection{Phase estimation}
To demonstrate the enhanced precision of our engineered mixed probe in quantum parameter estimation, we apply it to a typical quantum metrology application---quantum phase estimation \cite{qpe1,qpe2,qpe3}. While our earlier results indicate that the engineered probe state via variational optimization exhibits a significantly improved QFI, approaching the ultimate QCRB given by the optimized QFI necessitates an optimal or near-optimal readout protocol. Here we adopt an easily implementable measurement protocol known as the time-reversal-based readout (TRBR) protocol, which exploits time-reversal dynamics to disentangle probe states for feasible readout and has previously demonstrated on diverse platforms such as cold-atom cavity-QED systems \cite{time_re1}, Bose-Einstein condensates \cite{Linnemann2016}, and trapped ions \cite{Gilmore2021}. In our experiment, we employ the previously optimized state $\rho_\text{f}$ as the probe and encode the parameter to be estimated $\alpha$, i.e., $\rho_\text{f}^{\alpha}=e^{-i\alpha G}\rho_\text{f}e^{i\alpha G}$. To implement the TRBR protocol, we apply the inverse evolution $U_{\text{E}}^\dagger(\vec\theta)$ before projecting onto the initial state $\rho_0$, which is equivalent to applying a near-optimal measurement $\mathcal O_\text{rev}=U_{\text{E}}(\vec\theta)\rho_0U_{\text{E}}^\dagger(\vec\theta)$ on $\rho_\text{f}^\alpha$. Finally, we assess the performance of the optimized probe under the TRBR protocol according to the error propagation formula \cite{advanceQM}
	\begin{eqnarray}\label{VarC}
(\Delta\alpha_\text{f})^2=\frac{(\Delta\mathcal O_\text{rev})^2}{(d\langle\mathcal O_\text{rev}\rangle/d\alpha)^2},
	\end{eqnarray}
where $(\Delta\mathcal O_\text{rev})^2=\langle\mathcal O_\text{rev}^2\rangle-\langle\mathcal O_\text{rev}\rangle^2$ represents the quantum fluctuation of $\mathcal O_\text{rev}$, $\langle \mathcal O_\text{rev}\rangle=\text{Tr}(\rho_\text{f}^{\alpha}\mathcal O_\text{rev})$. Details of the experimental extraction of $(\Delta\alpha_\text{f})^2$ are elaborated in Methods.

	
 We benchmark the precision of the optimized mixed state $\rho_\text{f}$ against its classical counterpart $\rho_\text{cl}$, where $\rho_\text{cl}$ is generated by local operations on individual spins from $\rho_0$ with a SQL-like precision scaling $\Delta\alpha_\text{cl}\sim1/\sqrt N$\cite{MixedPRX}. The experimental result of $(\Delta\alpha_\text{f})^2/(\Delta\alpha_\text{cl})^2$ is depicted in Fig.~\ref{fig3} with red stars, closely matching the theoretical prediction indicated by the blue dashed line. Under the experimental condition $\epsilon\sim10^{-5}$, we have $\Delta\alpha_\text{cl}\sim1.7\times10^4$, and the optimum of $\Delta\alpha_\text{f}$ occurs around $\widetilde\alpha=0.08\pi$, being $\Delta\alpha_\text{f}\sim4.0\times10^{3}$. This results in a precision ratio $(\Delta\alpha_\text{f})^2/(\Delta\alpha_\text{cl})^2=0.056$, corresponding to a 12.4 dB improvement. This improvement in precision, greater than $\sqrt N$, is attributed to the complex eigen spectrum of mixed states, as discussed in Ref. \cite{MixedPRX} and further detailed in our Supplemental Material \cite{supp}. In practical implementation, we can asymptotically approach this local precision by adaptively adjusting $\alpha$ near $\widetilde\alpha$ with an additional control field \cite{Hentschel2011}. The QCRB is also plotted with a black solid line. The inset clarifies that while the TRBR protocol is suboptimal, the current engineered probe can still achieve quantum-enhanced precision close to the QCRB.

\section{CONCLUSIONS}

To summarize, we propose a novel scheme for variational quantum metrology with LE. We demonstrate its feasibility by engineering an optimal 10-spin mixed probe state on an NMR system, where the QFI is efficiently estimated using LE to guide the variational optimization. By utilizing the proposed time-reversal-based readout protocol, the engineered probe achieves a quantum-enhanced precision that approaches the optimal Quantum Cramer-Rao bound.

 The proposed variational scheme features several advantages for experimental implementation. First, since the measured Loschmidt echo provides a faithful lower bound for QFI \cite{cerezo2021}, this scheme can be extended to quantum systems with different purity. In addition, the scheme does not require detailed knowledge of the encoding dynamics during optimization, which is often unknown in practice. Furthermore, by utilizing VQCs designed with hardware-efficient Ans\"{a}tze and a measurement overhead that scales linearly with system size, our scheme demonstrates practical efficiency and scalability for extracting QFI. This work paves the way for broader applications of variational quantum metrology to diverse quantum sensing tasks and quantum systems. Future work could explore the use of gradient-based classical optimizers to enhance efficiency. For example, the parameter-shift rule \cite{Schuld2019} enables direct gradient evaluation on quantum processors, potentially improving optimization in complex parameter spaces. We also anticipate that future research will explore our scheme on various NISQ computers \cite{bluvstein2024,evered2023,guo2024}, demonstrating quantum-enhanced precision.
	\section{FUNDING}
\begin{acknowledgements}
	This work was supported by the Innovation Program for Quantum Science and Technology (2021ZD0303205), the National Natural Science Foundation of China (12261160569, 12404554), New Cornerstone Science Foundation through the XPLORER PRIZE, the Research Grants Council of Hong Kong (14309223, 14309624, 14309022), the Innovation Program for Quantum Science and Technology (2023ZD0300600), the Guangdong Provincial Quantum Science Strategic Initiative (GDZX2303007), China Postdoctoral Science Foundation (2024M762114) and Postdoctoral Fellowship Program of CPSF (GZC20231727).
\end{acknowledgements}

\section{METHODS}
\subsection{The connection between Loschmidt echo and QFI\label{app1}}

For the pure initial probe state $|\Psi_0\rangle$, the Loschmidt echo (LE) under engineering operation $U_\text{E}$ and encoding dynamics $G$ can be expressed as
\begin{eqnarray}\label{Eq:LEsupp}
	\mathcal{L}_\delta=|\langle\Psi_0|U_\text{E}^\dagger e^{-i\delta G}U_\text{E}|\Psi_0\rangle|^2.
\end{eqnarray}
By expanding Eq.(\ref{Eq:LEsupp}) with Taylor series around $\delta=0$, we have 
\begin{eqnarray}\notag
	\begin{aligned}
	\mathcal{L}_\delta=&\langle\Psi_f|e^{-i\delta G}|\Psi_f\rangle\langle\Psi_f|e^{i\delta G}|\Psi_f\rangle\\
	=&\left(1-i\delta\langle G\rangle-\frac{\delta^2}{2}\langle G^2\rangle+\frac{i\delta^3}{6}\langle G^3\rangle\right)\times\\
	&\left(1+i\delta\langle G\rangle-\frac{\delta^2}{2}\langle G^2\rangle-\frac{i\delta^3}{6}\langle G^3\rangle\right)+O(\delta^4)\\
	=&1-\delta^2\left(\langle G^2\rangle-\langle G\rangle^2\right)+O(\delta^4),
	\end{aligned}
\end{eqnarray}
where $\langle\cdot\rangle\equiv\langle\Psi_f|\cdot|\Psi_f\rangle$ and $|\Psi_f\rangle\equiv U_\text{E}|\Psi_0\rangle$. As the quantum Fisher information (QFI) for pure state is
\begin{eqnarray}
\mathcal F(|\Psi_f\rangle)=4\left(\langle G^2\rangle-\langle G\rangle^2\right),
\end{eqnarray}
we have \cite{QFILE}
\begin{eqnarray}
	\mathcal F\left(|\Psi_f\rangle\right)=\lim_{\delta\to0}4\frac{1-\mathcal{L}_{\delta}}{\delta^2}.
\end{eqnarray}
For the mixed engineered probe $\rho_\text{f}=U_\text{E}\rho_0U_\text{E}^\dagger$ with the eigendecomposition $\sum_{i=1}^d\lambda_i|\psi_i\rangle\langle\psi_i|$ and $d$ as the dimension of the Hilbert space, the LE can be computed as
\begin{eqnarray}\nonumber
	\begin{aligned}
		\mathcal{L}_\delta=&\text{Tr}\left(\rho_\text{f} e^{-i\delta G}\rho_\text{f} e^{i\delta G}\right)\\
		=&\text{Tr}\left(\sum_{i=1}\lambda_i|\psi_i\rangle\langle\psi_i| e^{-i\delta G}\sum_{j=1}\lambda_j|\psi_j\rangle\langle\psi_j| e^{i\delta G}\right)\\
		=&\sum_k\langle\psi_k|\sum_{i=1}\lambda_i|\psi_i\rangle\langle\psi_i| e^{-i\delta G}\sum_{j=1}\lambda_j|\psi_j\rangle\\
		&\times\langle\psi_j| e^{i\delta G}|\psi_k\rangle\\
		=&\sum_i\lambda_i^2-\delta^2\left(\sum_{i,j}\lambda_i\lambda_j|\langle\psi_i|G|\psi_j|^2\right.\\
		&\left.+\sum_i\lambda_i^2\langle\psi_i| G^2|\psi_i\rangle\right)+O(\delta^4)
	\end{aligned}
\end{eqnarray}
The zeroth-order term in the perturbation expansion, i.e., $\sum_i\lambda_i^2$, represents the purity of $\rho_0$, and it does not change under unitary transformation. While for the second-order terms, note that
\begin{eqnarray}\nonumber
	\begin{aligned}
	&\sum_i\lambda_i^2\langle\psi_i|G^2|\psi_i\rangle\\
	=&\frac{1}{2}(\sum_i\lambda_i^2\langle\psi_i|G^2|\psi_i\rangle
	+\sum_j\lambda_j^2\langle\psi_j|G^2|\psi_j\rangle)\\
	&\langle\psi_i|G^2|\psi_i\rangle =\langle\psi_i|G\sum_j|\psi_j\rangle\langle\psi_j|G|\psi_i\rangle\\
	=&\sum_j|\langle\psi_i|G|\psi_j\rangle|^2,
	\end{aligned}
\end{eqnarray}
we thus have
\begin{eqnarray}\nonumber
	\begin{aligned}
\mathcal{L}_\delta=&\sum_i\lambda_i^2+\sum_{i,j}\lambda_i\lambda_j\delta^2|\langle\psi_i|G|\psi_j\rangle|^2\\
&-\frac{\delta^2}{2}\sum_{i,j}\lambda_i^2\delta^2|\langle\psi_i|G|\psi_j\rangle|^2\\
&-\frac{\delta^2}{2}\sum_{i,j}\lambda_j^2\delta^2|\langle\psi_i|G|\psi_j\rangle|^2+O(\delta^4)\\
=&\sum_i\lambda_i^2-\frac{\delta^2}{4}\left(2\sum_{i,j}\left(\lambda_i-\lambda_j\right)^2|\langle\psi_i|G|\psi_j\rangle|^2\right).\\
\end{aligned}
\end{eqnarray}
Comparing with the QFI for mixed states
\begin{eqnarray}\label{mixFq}
 \mathcal{F}\left(\rho_\text{f}\right)=2 \sum_{i, j} \frac{\left(\lambda_{i}-\lambda_{j}\right)^{2}}{\lambda_{i}+\lambda_{j}}\left|\left\langle \psi_{i}\left|G\right| \psi_{j}\right\rangle\right|^{2},
\end{eqnarray}
we have
\begin{eqnarray}
	\mathcal F(\rho_\text{f})\ge\lim_{\delta\to0}4\frac{\Gamma(\rho_\text{f})-\mathcal{L}_\delta}{\delta^2},
\end{eqnarray}
where we used the fact that $\lambda_i+\lambda_j\le1$, and $\Gamma(\cdot)$ denotes the purity of the state. For highly mixed states where the eigenvalues are almost degenerate, i.e., $\lambda_i\approx\frac{1}{d}$ for $1\le i\le d$, we have 
\begin{eqnarray}\label{FqLE}
	\mathcal F(\rho_\text{f})\approx\lim_{\delta\to0}2d\frac{\Gamma(\rho_\text{f})-\mathcal{L}_\delta}{\delta^2},
\end{eqnarray}

\subsection{Experimental calibration of the precision of phase estimation}

$\Delta\alpha_\text{f}$ can be calibrated according to Eq. \eqref{VarC}, in which
\begin{eqnarray}\nonumber
	\begin{aligned}
		\langle{\mathcal O}_\text{rev}\rangle=&\text{Tr}\left(e^{-i\alpha G}\rho_\text{f} e^{i\alpha G}\mathcal O_\text{rev}\right)\\
		=&\frac{1}{2^{N}}+\frac{\epsilon^2}{2^{2N}}\text{Tr}(e^{-i\alpha G}\rho_\text{f}^\Delta e^{i\alpha G}\rho_\text{f}^\Delta),\\
		\langle{\mathcal O}_\text{rev}^2\rangle=&\text{Tr}\left(e^{-i\alpha G}\rho_\text{f} e^{i\alpha G}\mathcal O_\text{rev}^2\right)\\
		=&\frac{1}{2^{2N}}+\frac{\epsilon^2}{2^{2N+2}}\sum_i\gamma^2_i\\
		&+\frac{\epsilon^2}{2^{3N-1}}\text{Tr}(e^{-i\alpha G}\rho_\text{f}^\Delta e^{i\alpha G}\rho_\text{f}^\Delta)+O(\epsilon^3).
	\end{aligned}
\end{eqnarray}
In experiment, following the method in Refs. \cite{Girolami2014,Chen2021}, we extract $(\Delta\mathcal O_\text{rev})^2=\langle\mathcal O_\text{rev}^2\rangle-\langle\mathcal O_\text{rev}\rangle^2$ by substituting the experimental signal of $\text{Tr}(e^{-i\alpha G}\rho_\text{f}^\Delta e^{i\alpha G}\rho_\text{f}^\Delta)$ into equations above with $N,\gamma_i,\epsilon$ being known. The derivation $d\langle\mathcal O_\text{rev}\rangle/d\alpha$ is approximated with the finite difference approach
\begin{eqnarray}
	\frac{\langle\mathcal O_\text{rev}\rangle}{d\alpha}\approx \frac{\langle\mathcal O_\text{rev}\rangle_{\alpha+\delta^\prime}-\langle\mathcal O_\text{rev}\rangle_{\alpha-\delta^\prime}}{2\delta^\prime},
\end{eqnarray}
with $\delta^\prime=\frac{\pi}{50}$. For the experimental condition of $\epsilon\sim10^{-5}$, we have the precision of engineered state $\Delta\alpha_{\rm f}\sim4.0\times10^3$ at $\widetilde\alpha=0.08\pi$.

\end{document}


\title{Variational Quantum Metrology with Loschmidt Echo: Supplemental Material}

	\date{\today}
	\author{Ran Liu}
	\thanks{These authors contribute equally}
	\affiliation{CAS Key Laboratory of Microscale Magnetic Resonance and School of Physical Sciences, University of Science and Technology of China, Hefei 230026, China}
	\affiliation{CAS Center for Excellence in Quantum Information and Quantum Physics, University of Science and Technology of China, Hefei 230026, China}
	
	\affiliation{Institute of Quantum Precision Measurement, State Key Laboratory of Radio Frequency Heterogeneous Integration, College of Physics and Optoelectronic Engineering, Shenzhen University, Shenzhen 518060, China}
	\author{Ze Wu}
	\thanks{These authors contribute equally}
	\affiliation{CAS Key Laboratory of Microscale Magnetic Resonance and School of Physical Sciences, University of Science and Technology of China, Hefei 230026, China}
	\affiliation{CAS Center for Excellence in Quantum Information and Quantum Physics, University of Science and Technology of China, Hefei 230026, China}
	\author{Xiaodong Yang}
	\affiliation{Institute of Quantum Precision Measurement, State Key Laboratory of Radio Frequency Heterogeneous Integration, College of Physics and Optoelectronic Engineering, Shenzhen University, Shenzhen 518060, China}	
	\affiliation{Quantum Science Center of Guangdong-Hong Kong-Macao Greater Bay Area (Guangdong), Shenzhen 518045, China}
	\author{Yuchen Li}
	\affiliation{CAS Key Laboratory of Microscale Magnetic Resonance and School of Physical Sciences, University of Science and Technology of China, Hefei 230026, China}
	\affiliation{CAS Center for Excellence in Quantum Information and Quantum Physics, University of Science and Technology of China, Hefei 230026, China}
	\author{Hui Zhou}
	\affiliation{School of Physics, Hefei University of Technology, Hefei, Anhui 230009, China}
	\author{Zhaokai Li}
	\affiliation{CAS Key Laboratory of Microscale Magnetic Resonance and School of Physical Sciences, University of Science and Technology of China, Hefei 230026, China}
	\affiliation{CAS Center for Excellence in Quantum Information and Quantum Physics, University of Science and Technology of China, Hefei 230026, China}
	\affiliation{Hefei National Laboratory, University of Science and Technology of China, Hefei 230088, China}
	\author{Yuquan Chen}
	\affiliation{CAS Key Laboratory of Microscale Magnetic Resonance and School of Physical Sciences, University of Science and Technology of China, Hefei 230026, China}
	\affiliation{CAS Center for Excellence in Quantum Information and Quantum Physics, University of Science and Technology of China, Hefei 230026, China}
	
	\author{Haidong Yuan}
	\email{hdyuan@mae.cuhk.edu.hk}
	\affiliation{Department of Mechanical and Automation Engineering, The Chinese University of Hong Kong, Shatin, Hong Kong SAR, China}
	\author{Xinhua Peng}
	\email{xhpeng@ustc.edu.cn}
	\affiliation{CAS Key Laboratory of Microscale Magnetic Resonance and School of Physical Sciences, University of Science and Technology of China, Hefei 230026, China}
	\affiliation{CAS Center for Excellence in Quantum Information and Quantum Physics, University of Science and Technology of China, Hefei 230026, China}
	\affiliation{Hefei National Laboratory, University of Science and Technology of China, Hefei 230088, China}

	\maketitle

\section{Optimal quench time under encoding dynamics}\label{nois}
\begin{figure*}
		\includegraphics[scale=1]{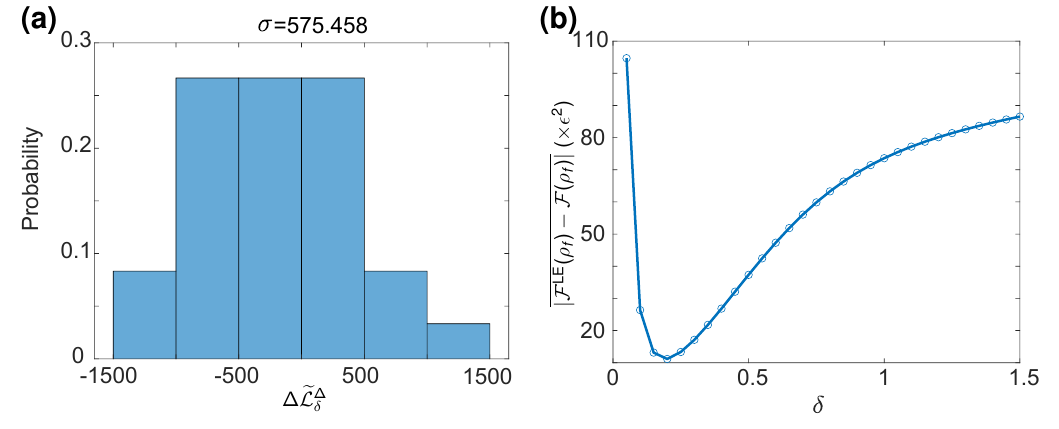} 
		\caption{(a) Calibration of experimental error for measuring $\mathcal{L}_\delta^\Delta$. (b) Numerical simulation of deviations on QFI under different quench time $\delta$. When $\delta=0.2$, the error gets its minimum.} \label{Fig1}
\end{figure*}

When extracting QFI from LE according to 
\begin{eqnarray}\label{FqLE}
		\mathcal F\left[\rho_\text{f},G\right]\approx\lim_{\delta\to0}2d\frac{\Gamma(\rho_\text{f})-\mathcal L_\delta}{\delta^2},
	\end{eqnarray}
	there are three factors that can lead to the deviation from the theoretical QFI: 
1) the eigenvalues of $\rho_\text{f}$ are not exactly equal to $1/d$; 2) the higher-order (>2) terms are ignored in Eq. \eqref{FqLE} but they can be non-zero; 3) the experimental errors when measuring the LE. In the following, we give an analysis on how these factors affect the deviation and decide an optimal value of $\delta$ for the experimental extraction of the QFI.
\begin{itemize}
	\item[1)] In our experiments, the equilibrium state is $\rho_0=(\mathds{1}+\epsilon\rho_0^\Delta)/2^{N}$ with $\rho_0^\Delta=\gamma_{\text{P}}\sigma_z^{\text{P}}+\gamma_{\text{H}}\sum_{j=1}^9\sigma_{j,z}^{\text{H}}$. The eigenvalues can then be expressed as $\lambda_i=1/2^{N}+\lambda_i^\Delta$ with $\lambda_i^\Delta\sim N\epsilon/2^{N}$. The error caused by taking $\lambda_i$ as $1/d=1/2^{N}$ can thus be obtained as 
	\begin{eqnarray}
		\begin{aligned}
		\mathcal E_1=&2\sum_{i,j}\left(\lambda_i-\lambda_j\right)^2|\langle\psi_i|G|\psi_j\rangle|^2\left|\frac{1}{\frac{2}{2^{N}}}-\frac{1}{\frac{2}{2^{N}}+\lambda_i^{\Delta}+\lambda_j^{\Delta}}\right|\\
		=&2\sum_{i,j}\left(\lambda^\Delta_i-\lambda^\Delta_j\right)^2|\langle\psi_i|G|\psi_j\rangle|^2\left|\frac{\lambda_i^{\Delta}+\lambda_j^{\Delta}}{\frac{2}{2^{N}}\left(\frac{2}{2^{N}}+\lambda_i^{\Delta}+\lambda_j^{\Delta}\right)}\right|\\
		\sim&\frac{(N\epsilon)^3}{2^{N+2}}
		\end{aligned}
	\end{eqnarray}

	\item[2)] For the Taylor series expansion of $\mathcal{L}_\delta$, its higher-order terms are non-negligible under finite $\delta$. The leading term, i.e., the fourth-order term, is
	\begin{eqnarray}
		\delta^4\left(\frac{1}{4}\sum_{i,j}\lambda_i\lambda_j|\langle\psi_i |G^2|\psi_j\rangle|^2+\frac{1}{12}\sum_i\lambda_i^2\langle\psi_i|G^4|\psi_i\rangle-\frac{1}{3}\sum_{i,j}\lambda_i\lambda_j\text{real}\left(\langle\psi_j|G|\psi_i\rangle\langle\psi_i|G^3|\psi_j\rangle\right)\right).
	\end{eqnarray}
	Consequently, it leads to the error on QFI
	\begin{eqnarray}\label{eq:E2}
		\mathcal E_2\sim\frac{\delta^2}{2^{N-1}}.
	\end{eqnarray}

\item[3)] As we mentioned in the main text, $\mathcal{L}_\delta$ is experimentally obtained by measuring its deviation part $\mathcal{L}_\delta^{\Delta}$, i.e., $\mathcal{L}_\delta=1/2^N+\epsilon^2\mathcal{L}_\delta^{\Delta}$. The deviations of the experimental results $\mathcal{L}_\delta^{\Delta,\text{exp}}$ from the theoretical one leads to the error on QFI,
\begin{eqnarray}\label{eq:E3}
	\mathcal E_3\sim\frac{2^{N+1}\epsilon^2\Delta\mathcal{L}^{\Delta}_\delta}{\delta^2}
\end{eqnarray}
with $\Delta\mathcal{L}^{\Delta}_\delta=\mathcal{L}^{\Delta,\text{exp}}_\delta-\mathcal{L}^{\Delta,\text{theo}}_\delta$. We calibrate the deviations from experiments by measuring $\mathcal{L}^{\Delta,\text{exp}}_\delta$ under 80 different $U_\text{E}(\vec{\theta})$ and calculating their standard deviation as $\sigma=575.46$. The probability histogram is shown in Fig. \ref{Fig1} (a).   
\end{itemize}
Since $\epsilon\sim10^{-5}$ at the thermal polarization, $\mathcal E_1$ is much smaller than $\mathcal E_2$ and $\mathcal E_3$, thus negligible. The influences of $\delta$ on $\mathcal E_2$ and $\mathcal E_3$ are opposite, as Eq.(\ref{eq:E2}) shows $\mathcal E_2$ increases with $\delta$ while Eq.(\ref{eq:E3}) shows $\mathcal E_3$ decreases with $\delta$. Thus there exists a tradeoff and an optimal $\delta$ needs to be determined to minimize the total error. As shown in Fig. \ref{Fig1} (b), we numerically simulate the mean of the difference between the QFI obtained from the LE under experimental errors and the theoretical one, i.e., $\overline{|\mathcal F^\text{LE}(\rho_\text{f})-\mathcal F(\rho_\text{f})|}$, under 50 different $U_\text{E}(\vec{\theta})$ for each $\delta$. The experimental error is simulated by adding artificial fluctuations on $\mathcal{L}^{\Delta}_\delta$ obeying normal distribution with a standard deviation $\sigma$. 
From the simulation we can see that the error is minimal at $\delta=0.2$, which is taken as the experimental quench time.  
\section{Details of the Nelder-Mead algorithm}
\begin{figure*}
		\includegraphics[scale=0.8]{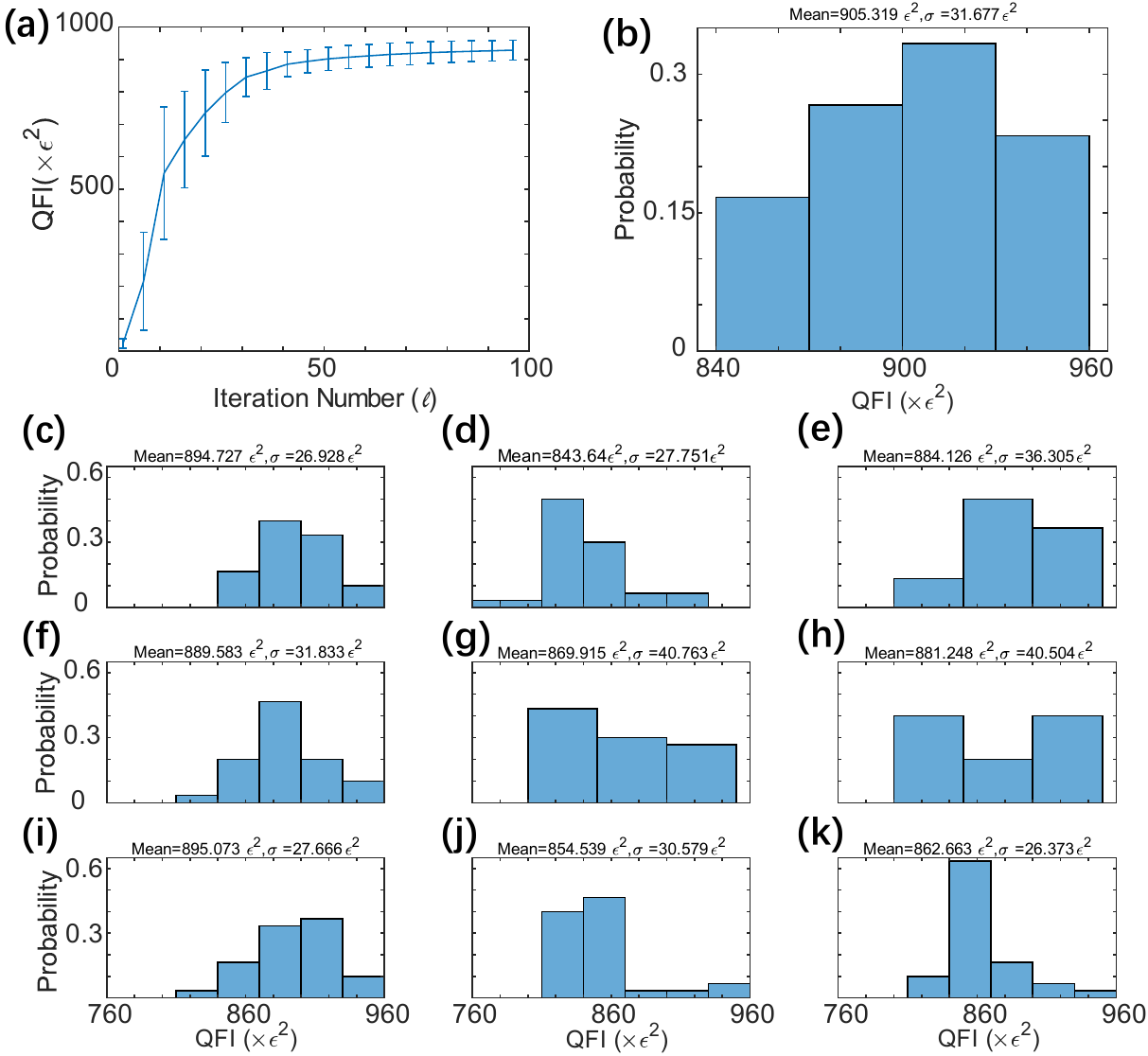} 
		\caption{Simulated iterations with NM algorithm. (a) Statistical result of 30 rounds of iteration under different initial simplexes and random fluctuation. When $l\ge70$, the optimization converges and approaches to its optimum. Statistical result of optimized QFI obtained from 100 rounds of iteration with (b) modified initial simplex and (c-k) random initial simplex. The modified one shows higher expectation to approach a large QFI.}\label{figNM} 
\end{figure*}

The Nelder-Mead algorithm provides a useful procedure for searching the minimum of a given function without derivatives. By rescaling a simplex consisting of $n+1$ vertices iteratively, this algorithm attempts to repalce the worst vertex by a better one. Here each vertex represents a sequence of $n$ parameters that can be tuned in the parametrized quantum circuit(PQC). The procedure of the Nelder-Mead algorithm is described as below.
\begin{itemize}
	\item[1.]\textbf{Ording}: Calculate the cost function of $n+1$ initial vertices and sort them as
	\begin{eqnarray}
		f(\vec{\theta}^{(1)})\le f(\vec{\theta}^{(2)})\le\cdots\le f(\vec{\theta}^{(n+1)}).
	\end{eqnarray}
where the cost function is LE with  $f(\vec{\theta}^{(i)})=\mathcal{L}_\delta(\vec{\theta}^{(i)})$.	For the 3-layer PQC used in the experiment we have $n=6$. 
	\item[2.]\textbf{Centroid}: Evaluate the cost function of the centroid of the best $n$ points, $f(\vec{\theta}^\text{ave})$, here $\vec{\theta}^\text{ave}=\sum_{i=1}^n\vec{\theta}^{(i)}/n$. 
	\item[3.]\textbf{transformation}: Replace the worst vertex $\vec{\theta}^{(n+1)}$ and the corresponding cost function with a better one by using reflection, expansion, contraction or shrink. The concrete rules for this transformation are as follows:
	\item[(1)]\textbf{Reflect}: Calculate the cost function of $f_r=f(\vec{\theta}^r)$ with $\vec{\theta}^r:=\vec{\theta}^\text{ave}+\alpha(\vec{\theta}^{(n+1)}-\vec{\theta}^{(n)})$ as the reflection point. $\alpha$ is the reflection coefficient and set as 1. If $f_1\le f_r\le f_n$, accept $\vec{\theta}^r$.
	
	\item[(2)]\textbf{Expand}: If $f_r<f_1$, calculate $f_e=f(\vec{\theta}^e)$ with $\vec{\theta}^e:=\vec{\theta}^\text{ave}+\gamma\cdot\alpha(\vec{\theta}^{(n+1)}-\vec{\theta}^{(n)})$ as the expansion point and $\gamma$ is the expansion coefficient and set as 2. If $f_e<f_r$ accept $\vec{\theta}^e$. Otherwise, accept $\vec{\theta}^r$.
	\item[(3)]\textbf{Contract}: (3a) If $f_n\le f_r\le f_{n+1}$, calculate $f_c:=f(\vec{\theta}^c)$ with $\vec{\theta}^c:=\vec{\theta}^\text{ave}+\beta\cdot\alpha(\vec{\theta}^{(n+1)}-\vec{\theta}^{(n)})$ as the outside contraction point and $\beta$ is the contraction coefficient and set as $0.5$. If $f_c\le f_r$, accept $f_c$. Otherwise, perform a shrink transformation. (3b) If $f_r\ge f_{n+1}$, calclulate $f_c:=f(\vec{\theta}^c)$ with $\vec{\theta}^c:=\vec{\theta}^\text{ave}-\beta\cdot\alpha(\vec{\theta}^{(n+1)}-\vec{\theta}^{(n)})$ as the inside contraction point and $\beta$ is set as $0.5$. If $f_c\le f_r$, accept $f_c$. Otherwise, perform a shrink transformation.
	\item[(4)]\textbf{Shrink}: Calculate $f_i:=f(\vec{\theta}^{(i)})$ with $\vec{\theta}^{(i)}:=\vec{\theta}^{(1)}+(1-\delta)\vec{\theta}^{(i)}$ and $i=2,3,...,n+1$. $\delta$ is the shrinkage coefficient and set as 0.5.
	\item[4.]\textbf{Termination tests}: If the result satisfies the stopping condition, terminate the iterations. Otherwise, change the iteration number as $l=l+1$ and continue at \textbf{Ording}.
\end{itemize}
Here the stopping condition is whether the optimized QFI is close to the optimum. We simulate 30 rounds of experimental iterations, and in each round the initial vertices ($l=1$) is generated randomly with $\vec\theta_i\in[0,2\pi]$ and $i=1,2,...,n+1$. To take the experimental errors into account, we add fluctuations on theoretical $\mathcal{L}^{\Delta}_\delta$ with the standard deviation $\sigma$ as obtained in Sec. \ref{nois}. The simulated result in Fig. \ref{figNM} (a) shows that the optimization converges and approaches to its optimum when $l\ge70$. Consequently, we stop the experimental iterations when $l=70$.

To improve the efficiency of the optimization, we further modify the initial simplex ($l=1$) as \cite{NM} 
\begin{eqnarray}
	\theta^{(i)}_j=\begin{cases}\theta^{(1)}_j+2\pi(\sqrt{n+1}-1) & i \neq j+1 \& \& i>1 \\ \theta^{(1)}_j+2\pi(\sqrt{n+1}+n-1) & i=j+1 \& \& i>1\end{cases}
\end{eqnarray}
and $\vec{\theta}^{(1)}=(0,0,...0)^\text{T}$. We simulate 100 rounds of optimization under this initial simplex and stop the iteration when $l=70$. The statistical result of optimized QFI is show in Fig. \ref{figNM} (b). Compared with the case of randomly generated initial simplex, as shown in Fig. \ref{figNM} (c)-(k), the modified one shows higher expectation to approach a large QFI under noise, thus adopted in the experiment.

\section{Precision bound of mixed-state quantum metrology}\label{secIV}

We study the achievable precision using initially mixed probe states, where the initial state can be transformed with unitary operations which does not alter its spectrum. 

For a quantum system initialized as a pure state for the estimation of $\theta$ in the dynamics $e^{-i\theta H}$, the optimal probe state is given by $|\psi_\text{Opt}\rangle=(|h_\text{max} +e^{i\varphi}|h_\text{min})/\sqrt2$, where $|h_\text{max,min}\rangle$ are the eigenvectors of $H$ corresponding to the maximal and minimal eigenvalues, and $\varphi$ is an arbitrary phase. In the case of $H=\sum_i\sigma_z^i/2$, $|\psi_{\rm{Opt}}\rangle$ corresponds to the NOON state and exhibits a 10lg($N$) dB enhancement in precision over SQL. However, for mixed initial state the situation is more complicated. For a $d$-dimensional mixed initial state, $\rho=\sum_{k=1}^d p_k\left|\psi_k\right\rangle\left\langle\psi_k\right|$, to estimate $\theta$ in the dynamics $e^{-i\theta H}$ with $H=\sum_{k=1}^dh_k|h_k\rangle\langle h_k|$, where $p_1 \geq \cdots \geq p_d$ and $h_1 \geq \cdots \geq h_d$ are ordered, the optimal probe state that can be obtained from the initial probe state under unitary operation is given by\cite{MixedPRX} 
\begin{equation}\label{optmixed}
    \rho_\text{Opt}=\sum_{k=1}^d p_k\left|\phi_k\right\rangle\left\langle\phi_k\right|
\end{equation}
with
\begin{equation}
    \left|\phi_k\right\rangle= \begin{cases}\frac{\left|h_k\right\rangle+\left|h_{d-k+1}\right\rangle}{\sqrt{2}} & \text { if } 2 k<d+1 \\ \left|h_k\right\rangle & \text { if } 2 k=d+1 \\ \frac{\left|h_k\right\rangle-\left|h_{d-k+1}\right\rangle}{\sqrt{2}} & \text { if } 2 k>d+1.\end{cases}
\end{equation}\label{MaxFq}
The corresponding Quantum Fisher Information (QFI) is given by, 
\begin{equation}\label{maxFQ}
	\mathcal F_\text{Opt}=\frac{1}{2}\sum_{k=1}^dp_{k,d-k+1}(h_k-h_{d-k+1})^2
\end{equation}
with
\begin{equation}
	p_{k, l}= \begin{cases}0 & \text { if } p_k=p_{l}=0 \\ \frac{\left(p_k-p_{l}\right)^2}{p_k+p_{l}} & \text { else. }\end{cases}.
\end{equation}

To investigate the scaling of the precision bound for mixed states, we consider a specific $N$-qubit mixed-state system initialized as $\rho=\rho_i^{\otimes N}$, where each qubit is given by $\rho_i=(\mathds{1}+\epsilon\sigma_z)/2$, and $\epsilon\in[0,1]$ represents polarization. This state is similar to a thermal state in NMR experiments when $\epsilon\sim10^{-5}$. Under the encoding dynamics $H=\sum_i\sigma_z^i/2$, the classical uncorrelated probe state generated by optimal local unitary operations on individual spin is $\rho_{i,x}^{\otimes N}$, where $\rho_{i,x}=(\mathds{1}+\epsilon\sigma_x)/2$. The corresponding QFI of $\rho_{i,x}^{\otimes N}$ is $\mathcal F_{\rm{cl}}=\epsilon^2N$, thus a scaling of standard quantum limit (SQL) \cite{MixedPRX}. The QFI of the probe state generated by the optimal global (entangling) unitary operation is given by
\begin{equation}\label{eq:suppQFI}
	\mathcal F_\text{Opt}=\sum_{m=0}^{N-1} \frac{(N-2 m)^2\left[(1+\epsilon)^{N-m} (1-\epsilon)^m-(1+\epsilon)^m (1-\epsilon)^{N-m}\right]^2}{2^N\left[(1+\epsilon)^{N-m} (1-\epsilon)^m+(1+\epsilon)^m (1-\epsilon)^{N-m}\right]}\left(\begin{array}{c}
        N-1 \\
        m
        \end{array}\right) \geq  \epsilon^2N^2.
\end{equation}
The enhancement over the $F_{\rm{cl}}$ of classical uncorrelated state is thus greater than $N$.

In our experiment, the initial state of each spin is $\rho_i^\prime=(\mathds{1}+\epsilon\gamma_i\sigma_z)/2$, where $\gamma_i$ related to the gyromagnetic ratios of different nuclei. This is slightly different from the state $\rho_i^{\otimes N}$ with $\rho_i=(\mathds{1}+\epsilon\sigma_z)/2$ as we mentioned above, which is due to different gyromagnetic ratios in the molecule used in our experiment. For the experimental thermal polarization $\epsilon\sim10^{-5}$, we have $\bigotimes_{i=1}^N\rho_i^\prime\approx(\mathds{1}+\epsilon\rho_\text{eq}^\Delta)/2^{N}$ with $\rho_\text{eq}^\Delta=\sum_{j=1}^{N}\gamma_{j}\sigma_z^{j}$, i.e., the equilibrium state in our manuscript. Similarly, the classical uncorrelated state that can be generated from $\bigotimes_{i=1}^N\rho_i^\prime$ via optimal local unitary operation on individual spin is $\bigotimes_{i=1}^N\rho_{i,x}^\prime$ with $\rho_{i,x}^\prime=(\mathds{1}+\epsilon\gamma_i\sigma_x)/2$, whose QFI equals to $\epsilon^2\sum_{i}^N\gamma_i^2$. For the 10-spin TMP molecule in our experiment, we have $\gamma_1=0.8,\gamma_{2,3,...10}=2$, this gives the maximal QFI under the local unitary operation as $36\epsilon^2$, which serves as the classical precision bound for $N=10$. The QFI of the optimal state, which can be obtained from Eq. \eqref{maxFQ}, is $989\epsilon^2$, a 14.4 dB improvement over the classical case. This is different from maximal 10 dB improvement in the case of pure initial probe state.

\section{Structure of variational quantum circuit and its reachable set}
\begin{figure*}
	\includegraphics[scale=1]{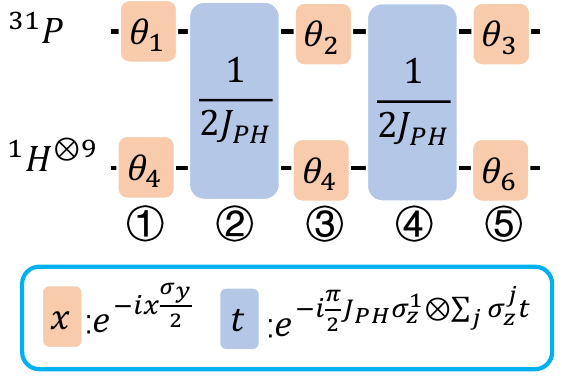} 
	\caption{The structure of PQC used to engineer quantum probe, which consists of single-qubit rotations and entangling gates. Its reachable set of quantum states can cover the optimal quantum probe when $\sin\theta_2=1,\sin\theta_6=1,\cos\theta_1\cos\theta_3\sin\theta_4=1$.}\label{sPQC}
\end{figure*}

The structure of variational quantum circuit (VQC) used to realize the engineering operation $U_\text{E}$ is shown in Fig. \ref{sPQC}. The PQC has 3 layers with the first layer being local rotations along $y$-axis while the others consisting of an entangling gate and local rotations. The entangling gate is realized by the free evolution under the system Hamiltonian $H_\text{NMR}=\frac{\pi}{2}J_\text{PH}\sigma_z^1\otimes\sum_{j=2}^{10}\sigma_z^j$ with the interval $\tau=1/2J_\text{PH}$, and the parameters are the angles of local rotation $\vec{\theta}:=(\theta_1,\theta_2,...\theta_6)^\text{T}$. We now show that the reachable set of quantum state generated by $U_\text{E}(\vec{\theta})$, denote as $\mathcal R_{U_\text{E}}$, includes the optimal probe state.  

The polarized part of initial equilibrium state can be written as  $\gamma_\text{P}Z(I^{\otimes9})/9+\gamma_\text{H}I(ZI^{\otimes8})$, here $X,Y,Z,I$ represent $\sigma_{x,y,z},\mathds{1}_2$ respectively, and $(\cdot)$ represent the summation over all the indistinguishable permutations, such as $(I^{\otimes9}):=\sum_{i=1}^9I^{\otimes9}\cdots,(ZI^{\otimes8}):=ZII\cdots+IZI\cdots+IIZ\cdots+...$ . The evolved state after $e^{-i\left(\theta_1\sigma_y/2+\theta_4\sum_{j=2}^{10}\sigma_{y}^z/2\right)}$, corresponding to stage $\textcircled{1}$ in Fig. \ref{sPQC}, is
\begin{eqnarray}
	\begin{aligned}
\textcircled{1}: &Z(I^{\otimes9}) \rightarrow \cos \theta_{1} Z(I^{\otimes9})+\sin \theta_{1} X(I^{\otimes9})\\
&I(ZI^{\otimes8}) \rightarrow \cos \theta_{4}I(ZI^{\otimes8})+\sin\theta_{4}I(XI^{\otimes8})\\
&\rho_{\textcircled{1}}=\gamma_\text{P}\left[\cos\theta_1Z(I^{\otimes9})+\sin\theta_1X(I^{\otimes9})\right]/9+\gamma_\text{H}\left[\cos \theta_{4}I(ZI^{\otimes8})+\sin\theta_{4}I(XI^{\otimes8})\right]
	\end{aligned}
\end{eqnarray}
Since $\left[\sigma_z^1\sigma_{z}^i,\sigma_z^1\sigma_{z}^j\right]=0$ for $i,j=1,2,...9$, the different terms in $H_{\text{NMR}}$ can then be applied in turn,
\begin{eqnarray}
	\begin{aligned}
\textcircled{2}: &Z(I^{\otimes9}) \rightarrow Z(I^{\otimes9})\\
&X(I^{\otimes9})\stackrel{I_{z}^{0} I_{z}^{1}}{\longrightarrow}9YZI^{\otimes8}\stackrel{I_{z}^{0} I_{z}^{2}}{\longrightarrow}-9XZZI^{\otimes7}\stackrel{I_{z}^{0} I_{z}^{3}}{\longrightarrow}\cdots\stackrel{I_{z}^{0} I_{z}^{9}}{\longrightarrow}Y(Z^{\otimes9})\\
&I(ZI^{\otimes8})\to I(ZI^{\otimes8})\\
&I(XI^{\otimes8})\to Z(YI^{\otimes8})\\
&\rho_{\textcircled{2}}=\gamma_\text{P}\left[\cos\theta_1Z(I^{\otimes9})+\sin\theta_1Y(Z^{\otimes9})\right]/9+\gamma_\text{H}\left[\cos \theta_{4}I(ZI^{\otimes8})+\sin\theta_{4}Z(YI^{\otimes8})\right].
	\end{aligned}
\end{eqnarray}
Similarly, we have 
\begin{eqnarray}
	\begin{aligned}
	\textcircled{3}:&Z(I^{\otimes9})\to\cos\theta_2Z(I^{\otimes9})+\sin\theta_2X(I^{\otimes9})\\
	&Y(Z^{\otimes9})\to\sum_{i=1}^{10}\cos^{10-i}\theta_5\sin^{i-1}\theta_5Y(Z^{\otimes10-i}X^{\otimes i-1})\\
	&I(ZI^{\otimes8})\to\cos\theta_5I(ZI^{\otimes8})+\sin\theta_5I(XI^{\otimes8})\\
	&Z(YI^{\otimes8})\to\cos\theta_2Z(YI^{\otimes8})+\sin\theta_2X(YI^{\otimes8})\\
	&\rho_{\textcircled{3}}=\gamma_\text{P}\left[\cos\theta_1(\cos\theta_2Z(I^{\otimes9})+\sin\theta_2X(I^{\otimes9}))+\sin\theta_1(\sum_{i=1}^{10}\cos^{10-i}\theta_5\sin^{i-1}\theta_5Y(Z^{\otimes10-i}X^{\otimes i-1}))\right]/9\\
&+\gamma_\text{H}\left[\cos\theta_{4}(\cos\theta_5I(ZI^{\otimes8})+\sin\theta_5I(XI^{\otimes8}))+\sin\theta_{4}(\cos\theta_2Z(YI^{\otimes8})+\sin\theta_2X(YI^{\otimes8}))\right]\\
	\end{aligned}
\end{eqnarray}
\begin{eqnarray}
	\begin{aligned}
\textcircled{4}:&Z(I^{\otimes9})\to Z(I^{\otimes9})\\
&X(I^{\otimes9})\to Y(Z^{\otimes9})\\
&Y(Z^{\otimes9})\stackrel{I_{z}^{1} I_{z}^{2}}{\longrightarrow}-9XIZ^{\otimes8}\stackrel{I_{z}^{1} I_{z}^{3}}{\longrightarrow}-9YIIZ^{\otimes7}\stackrel{I_{z}^{1} I_{z}^{4}}{\longrightarrow}\cdots\stackrel{I_{z}^{1} I_{z}^{10}}{\longrightarrow}X(I^{\otimes9})\\
&Y(XZ^{\otimes8})\to Y(XI^{\otimes8})\\
&\text{due to }YXZ^{\otimes8}\stackrel{I_{z}^{1} I_{z}^{2}}{\longrightarrow}YXZ^{\otimes8}\stackrel{I_{z}^{1} I_{z}^{3}}{\longrightarrow}XXIZ^{\otimes7}\stackrel{I_{z}^{1} I_{z}^{4}}{\longrightarrow}\cdots\stackrel{I_{z}^{1} I_{z}^{10}}{\longrightarrow}YXI^{\otimes8}\\
&\cdots\\
&Y(ZX^{\otimes8})\to X(IX^{\otimes8})\\
&\text{due to }YZX^{\otimes8}\stackrel{I_{z}^{1} I_{z}^{2}}{\longrightarrow}XIX^{\otimes8}\stackrel{I_{z}^{1} I_{z}^{3}}{\longrightarrow}XIX^{\otimes8}\stackrel{I_{z}^{1} I_{z}^{4}}{\longrightarrow}\cdots\stackrel{I_{z}^{1} I_{z}^{10}}{\longrightarrow}XIX^{\otimes8}\\
&Y(X^{\otimes9})\to Y(X^{\otimes9})\\
&I(ZI^{\otimes8})\to I(ZI^{\otimes8})\\
&I(XI^{\otimes8})\to Z(YI^{\otimes8})\\
&Z(YI^{\otimes8})\to I(XI^{\otimes8})\\
&X(YI^{\otimes8})\to X(YZ^{\otimes8})\\
&\text{due to }XYI^{\otimes8}\stackrel{I_{z}^{1} I_{z}^{2}}{\longrightarrow}XYI^{\otimes8}\stackrel{I_{z}^{1} I_{z}^{3}}{\longrightarrow}YYZI^{\otimes7}\stackrel{I_{z}^{1} I_{z}^{4}}{\longrightarrow}\cdots\stackrel{I_{z}^{1} I_{z}^{10}}{\longrightarrow}XYZ^{\otimes8}
	\end{aligned}
\end{eqnarray}
\begin{eqnarray}
	\begin{aligned}
&\rho_{\textcircled{4}}=\gamma_\text{P}\left[\cos\theta_1(\cos\theta_2Z(I^{\otimes9})+\sin\theta_2Y(Z^{\otimes9}))+\sin\theta_1\sum_{i=1}^{10}\cos^{10-i}\theta_5\sin^{i-1}\theta_5(X\text{or}Y)(I^{\otimes10-i}X^{\otimes i-1})\right]/9\\
&+\gamma_\text{H}\left[\cos\theta_{4}(\cos\theta_5I(ZI^{\otimes8})+\sin\theta_5Z(YI^{\otimes8}))+\sin\theta_{4}(\cos\theta_2I(XI^{\otimes8})+\sin\theta_2X(YZ^{\otimes8}))\right]\\
	\end{aligned}
\end{eqnarray}
\begin{eqnarray}
	\begin{aligned}
\textcircled{5}: &Z(I^{\otimes9})\to\cos\theta_3Z(I^{\otimes9})+\sin\theta_3X(I^{\otimes9})\\
&Y(Z^{\otimes9})\to\sum_{i=1}^{10}\cos^{10-i}\theta_6\sin^{i-1}\theta_6Y(Z^{\otimes10-i}X^{\otimes i-1})\\
&X(I^{\otimes9})\to\cos\theta_3X(I^{\otimes9})-\sin\theta_3Z(I^{\otimes9})\\
&Y(XI^{\otimes8})\to\cos\theta_6Y(XI^{\otimes9})-\sin\theta_6Y(ZI^{\otimes9})\\
&\cdots\\
&X(IX^{\otimes8})\stackrel{\theta_3I_y^1}{\longrightarrow}\cos\theta_3X(IX^{\otimes8})-\sin\theta_3Z(IX^{\otimes8})\\
&\stackrel{\theta_6\sum_{i}I_{y}^i}{\longrightarrow}\cos\theta_3(\sum_{i=1}^9\cos^{9-i}\sin^{i-1}\theta_6X(IX^{\otimes9-i}Z^{\otimes i-1}))-\sin\theta_3(\sum_{i=1}^9\cos^{9-i}\sin^{i-1}\theta_6Z(IX^{\otimes9-i}Z^{\otimes i-1}))\\
&Y(X^{\otimes9})\to\sum_{i=1}^{10}\cos^{10-i}\theta_6(-\sin\theta_6)^{i-1}Y(X^{\otimes10-i}Z^{\otimes i-1})\\
&I(ZI^{\otimes8})\to\cos\theta_6I(ZI^{\otimes8})+\sin\theta_6I(XI^{\otimes8})\\
&Z(YI^{\otimes8})\to\cos\theta_3Z(YI^{\otimes8})+\sin\theta_3X(YI^{\otimes8})\\
&I(XI^{\otimes8})\to\cos\theta_6I(XI^{\otimes8})-\sin\theta_6I(ZI^{\otimes8})\\
&X(YZ^{\otimes8})\stackrel{\theta_3I_y^1}{\longrightarrow}\cos\theta_3X(YZ^{\otimes8})-\sin\theta_3Z(YZ^{\otimes8})\\
&\stackrel{\theta_6\sum_{i}I_{y}^i}{\longrightarrow}\cos\theta_3(\sum_{i=1}^9\cos^{9-i}\sin^{i-1}\theta_6X(YZ^{\otimes9-i}X^{\otimes i-1}))-\sin\theta_3(\sum_{i=1}^9\cos^{9-i}\sin^{i-1}\theta_6Z(YZ^{\otimes9-i}X^{\otimes i-1}))\\
	\end{aligned}
\end{eqnarray}
Consequently, $\mathcal R_{U_\text{E}}$ can be given by $\textcircled{5}$ with $\theta_1,\theta_2,...\theta_6\in[0,2\pi]$.

The form of optimal probe is given by Eq. \eqref{optmixed}, and it guarantees that the amplitude of highest-order coherence is maximal, as the elements with higher-order coherence in probe state show faster phase accumulation. Under the encoding dynamics $G=\sum_{k=1}^{10}\sigma_z/2$, $X,Y$ contribute to the order of coherence while $I,Z$ do not. In the following, we consider the highest-order, i.e., 10-order, coherence in probe,
\begin{eqnarray}
	\rho^\Delta_{(10)}=\frac{\gamma_\text{P}}{9}\left[\cos\theta_1\sin\theta_2\sin^{9}\theta_6Y(X^{\otimes9})+\sin\theta_1\sin^9\theta_5\cos^{9}\theta_6Y(X^{\otimes9})\right]+\gamma_\text{H}\sin\theta_4\sin\theta_2\cos\theta_3\sin^8\theta_6X(YX^{\otimes8}).
\end{eqnarray}
When $\sin\theta_2=1,\sin\theta_6=1,\cos\theta_1\cos\theta_3\sin\theta_4=1$, the contriubtion of $\rho^\Delta_{(10)}$ is maximized and the corresponding probe is optimal. So $\mathcal R_{U_\text{E}}$ can cover the optimal probe.

\section{Experimental procedure for measuring LE and its scalability in quantum circuits}
\begin{figure*}
	\includegraphics[scale=1]{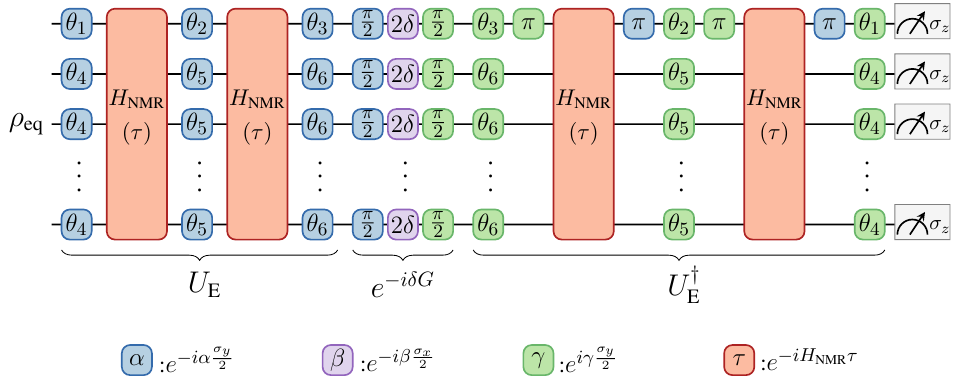} 
	\caption{Quantum circuit for measuring LE in the NMR experiment.}\label{QC_NMR}
\end{figure*}

The LE can be effectively constructed in the NMR experiments by performing a forward evolution $U_E$, a perturbation $e^{-i\delta G}$ with $\delta\to0$, and a reverse evolution $U_E^\dagger$ on the quantum system, and finally extracted by measuring the expectation of $\sigma_z$ of each spin. Specifically, the LE can be expressed as $\mathcal L_\delta=\text{Tr}\left[V_\delta(\vec{\theta})\rho_0V_\delta^\dagger(\vec{\theta})\rho_0\right]$, where $V_\delta(\vec{\theta})= U_\text{E}^\dagger(\vec{\theta}) e^{-i\delta G}U_\text{E}(\vec{\theta})$, and it can be experimentally extracted via the following procedures in the NMR experiments, as shown in Fig. \ref{QC_NMR}. 
 
  1). Start from the equilibrium state at room temperature of the NMR system $\rho_\text{eq}=(\mathds{1}+\epsilon\rho_\text{eq}^\Delta)/2^{N}$. Here, $\rho_\text{eq}^\Delta=\sum_{j=1}^{N}\gamma_{j}\sigma_z^{j}/2$, $\epsilon$ is the thermal polarization ($\sim10^{-5}$) and $\gamma_{j}$ is the relative gyromagnetic ratio of the corresponding nuclear. 
  
  2). Perform a forward evolution $U_\text{E}$, which consists of single-qubit rotations and free evolution under the system Hamiltonian $H_{\text{NMR}}=\frac{\pi}{2}J_{\text{PH}}\sigma_z^1\otimes\sum_{j=2}^{10}\sigma_z^j$ with a duration $\tau=1/2J_\text{PH}$.
  
  3). Perform the perturbation $e^{-i\delta G}$ with $G=\sum_{j=1}^N\sigma_z^j$. This is realized by single-qubit rotations with a small angle $2\delta$. 
  
  4). Perform the reverse evolution $U^\dagger_\text{E}$ by substituting each operation in $U_\text{E}$ with its inverse and applying them in reverse order. For single-qubit rotations, the reverse is achieved by adjusting the phase of the rotation. For free evolution under  $H_{\text{NMR}}$, the reverse is implemented by applying $\pi$ pulses to the first spin at both the beginning and the end of the free evolution.

  5). Project the evolved state onto the initial state $\rho_\text{eq}$. For the initial equilibrium state, this process can be equally expressed as $\mathcal L_\delta=c_0+\sum_{j=1}^Nc_j\text{Tr}\left[V_\delta(\vec{\theta})\rho_0V_\delta^\dagger(\vec{\theta})\sigma_z^j\right]$ with $c_j$ being constants. So in the NMR experiment we measure $\text{Tr}\left[V_\delta(\vec{\theta})\rho_0V_\delta^\dagger(\vec{\theta})\sigma_z^j\right]$, i.e., the $z$-direction polarizations of each spin. Obviously, the measurement overhead scales linearly with the size of our system. 
  
  Apart from the implementation in the 10-spin NMR experiment, we continue to analyze the scalability of measuring LE when $N$ is large and its feasibility in other quantum systems. Here, we employed the hardware-efficient Ans\"{a}tze for the design of variational quantum circuits $U_\text{E}$. These circuits are constructed from entangling layers $W_l$ and parameterized single-qubit rotation layers $U_l(\vec\theta_l)$, expressed as: 
\begin{eqnarray}
	U_E=\prod_{l=1}^L U_l\left(\boldsymbol{\theta}_l\right) W_l,
\end{eqnarray}
where $L$ denotes the number of layers. Moreover, these circuits employ a limited set of quantum gates, i.e., typically two-qubit entangling gates for $W_l$ and single-qubit gates for $U_l(\vec\theta_l)$. This ensures the efficient implementation of the inverse of each gate in the circuit. According to
\begin{eqnarray}
	U_E^\dagger=\prod_{l=1}^L W_{L-l+1}^\dagger U_{L-l+1}^\dagger\left(\boldsymbol{\theta}_{L-l+1}\right),
\end{eqnarray}
we can implement the reverse $U_E^\dagger$ by applying the reverse of each gate in $U_E$ in the reverse order. This ensures that the resource requirements for realizing $U_E^\dagger$ remain comparable to those for $U_E$, making the approach feasible when the size of system grows.

\begin{figure*}
	\includegraphics[scale=1]{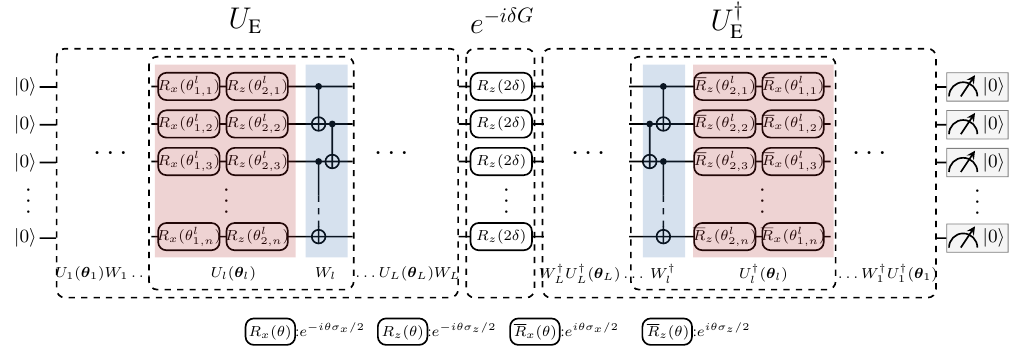} 
	\caption{Quantum circuit for measuring LE with the Hardware-efficient Ansatz containing CNOT gates.}\label{QC_circuit}
\end{figure*}
 This Ansatz also accommodates constraints such as limited qubit connectivity and restricted gate sets of quantum hardware, thus can be implemented on other quantum systems. In our experiment, the entangling gate is  $e^{-i\pi\sigma_z^1\sigma_z^i},(i>1)$ and can be reversed using  $\pi$-pulses. This design can be extended to other physical systems with scalable two-qubit gates, such as CNOT or CZ gates for superconducting qubits or XX gates for trapped ions. For example, the quantum circuit for measuring LE with the Hardware-efficient Ansatz containing CNOT gates can be implemented in Fig. \ref{QC_circuit}.

\section{Analysis of experimental errors and techniques for mitigating them}
As mentioned in the main text, the relative error of measured $\mathcal{L}_\delta^{\Delta}$ during the iteration is $1.35\%$. The sources of experimental errors include the pulse error, relaxation and measurement error. We employ different techniques to mitigate them. A detailed introduction of these techniques and the analysis of the experimental errors, aided with numerical simulations, are listed in the following.

\begin{itemize}
	\item[1.]\textbf{Pulse error}: The pulse error can lead to the deviation between the actually implemented rotation angles and the desired ones. This is mainly caused by the imperfect calibration of $\pi/2-$pulse in NMR experiment. To improve the robustness of the pulses, we replace the regular pulses with the BB1 sequence $R_{\phi}(\theta)\to R_{\phi}(\pi)R_{3\phi}(2\pi)R_{\phi}(\pi)R_{\phi}(\theta)$, where $\theta$ and $\phi$ is the target angle and phase, $\phi=\arccos(-\theta/4\pi)$. We simulate the pulse error by adding random fluctuation with 5$\%$ relative distortions to the amplitude of the pulses. The result shows that the relative error of $\mathcal{L}_\delta^{\Delta}$ caused by the pulse error decreases from $8.1\%$ to $4.5\times10^{-2}\%$ after applying the BB1 sequence. Thus the pulse error is significantly suppressed.
	
	\item[2.]\textbf{Relaxation}: 
	The quantum probe inevitably interacts with the environment during the evolution. The noise in NMR experiment can be described by the phase damping channel, $\mathcal{E}_{\text{PD}}$, and the generalized amplitude damping channel, $\mathcal{E}_{\text{GAD}}$. The effect of the phase damping channel on the density matrix $\rho$ can be approximately expressed as $\rho \rightarrow \mathcal{E}_{\text{PD}}^{N} \circ\cdots \circ\mathcal{E}_{\text{PD}}^{2} \circ \mathcal{E}_{\text{PD}}^{1}(\rho)$, where $\mathcal{E}_{\text{PD}}^{i}(\rho)=\left(1-\xi_{i}\right) \rho+\xi_{i} \sigma_{z}^i \rho \sigma_{z}^i$, $\xi_{i}=\frac{1}{2}\left[1-\exp \left(-\Delta t / T_{2}^{i}\right)\right]$ with $T_2^i$ as the transversal relaxation time of the $i$th spin. The influence of the generalized amplitude damping can be approximately characterized as $\rho \rightarrow \mathcal{E}_{\text{GAD}}^{N} \circ\cdots \circ\mathcal{E}_{\text{GAD}}^{2} \circ \mathcal{E}_{\text{GAD}}^{1}(\rho)$, where $\mathcal{E}_{\mathrm{GAD}}^{j}(\rho)=\sum_{s} E_{s}^{j} \rho E_{s}^{j^{\dagger}}$, 
\begin{eqnarray}\label{relaxd}
	\begin{aligned}
		&E_{1}^{j}=\sqrt{\frac{1}{2}}\left(\begin{array}{cc}
		1 & 0 \\
		0 & \sqrt{1-\eta_{j}}
		\end{array}\right), \quad E_{2}^{j}=\sqrt{\frac{1}{2}}\left(\begin{array}{cc}
		0 & 0 \\
		\sqrt{\eta_{j}} & 0
		\end{array}\right), \\
		&E_{3}^{j}=\sqrt{\frac{1}{2}}\left(\begin{array}{cc}
		\sqrt{1-\eta_{j}} & 0 \\
		0 & 1
		\end{array}\right), \quad E_{4}^{j}=\sqrt{\frac{1}{2}}\left(\begin{array}{cc}
		0 & \sqrt{\eta_{j}} \\
		0 & 0
		\end{array}\right),
	\end{aligned}
\end{eqnarray}
$\eta_{j}=1-\exp \left(-\Delta t / T_{1}^{j}\right)$ with $T_1^j$ as the longitudinal relaxation time of the $j$th qubit. The relaxation time for $^{31}$P nuclear spin and $^{1}$H nuclear spins are $T_2^{\text{P}}=1.30$ sec,$T_1^{\text{P}}=5$ sec and $T_2^{\text{H}}=1.26$ sec,$T_1^{\text{H}}=4.2$ sec, respectively. Moreover, a main source of phase damping in NMR experiment is the inhomogeneity of static magnetic field, and we suppress this effect by employing the refocusing sequences during the free evolution. The evolution time in a single experiment is 187 msec, thus the effect of the relaxation is not negligible. The numerical results show that the relative error of $\mathcal{L}_\delta^{\Delta}$ caused by the relaxation is $8.2\%$. To compensate for signal decay caused by relaxation during the evolution, we additionally measure a reference signal $S_0(\vec\theta)=Tr[V_0(\vec\theta)\rho V^\dagger_0(\vec\theta)\sigma_z]$ with $V_0(\vec\theta)=U_E^\dagger(\vec\theta)U_E(\vec\theta)$. Since the decay levels of  $S^\text{exp}(\vec{\theta})$  (experimental measurement of  $S(\vec{\theta})$ ) and  $S_0^\text{exp}(\vec{\theta})$  (experimental measurement of  $S_0(\vec{\theta})$ ) are approximately the same due to the similarity in evolutions  $V_\delta(\vec{\theta})$ and  $V_0(\vec{\theta})$, we use the following relation for calibration:
\[
 	\frac{S^\text{exp}(\vec\theta)}{S(\vec\theta)}\approx \frac{S^\text{exp}_0(\vec\theta)}{S_0(\vec\theta)}.
 	\]
  Note that $V_0(\vec\theta)=\mathds 1$, so $S_0(\vec\theta)$ does not depend on $\vec\theta$ and can be written as a known constant $S_0$. Consequently, we have the calibrated value for the target measurement $S(\vec\theta)$ as
  \[
 	S(\vec\theta)=\frac{S_0(\vec\theta)}{S^\text{exp}_0(\vec\theta)}\times S^\text{exp}(\vec\theta).
 \]
 Following the calibration procedure outlined above, we conducted numerical simulations based on the relaxation dynamics described in Eq.~\eqref{relaxd}, demonstrating that the relative error can be reduced to $0.98\%$.

\item[3.]\textbf{Measurement error}: The systematic errors of pulse and relaxation lead to consistent deviation from the theoretical expectation in the experiment. While the measurement error comes from the stochastic fluctuations on NMR signal and thus corresponds to the error bound of repetitive measurements. Its effect can be calibrated from the signal-to-noise ratio (SNR) of the NMR spectra. To improve the SNR of the $^{31}$P channel, protons are decoupled with composite pulses when measuring the signal of $^{31}\text{P}$ nuclear. According to the numerical simulation, the relative error of $\mathcal{L}_\delta^{\Delta}$ caused by the measurement error decreases from $2.5\times10^{-2}\%$ to $1.8\times10^{-2}\%$ after applying the decoupling operation. The absolute number of experimental error bounds can be analyzed via linear error propagation. For any linear function $f(x_1,x_2,...x_j,...)=\sum_jc_jx_j$, the variance is obtained from
\begin{equation}
	\sigma_f^2=\sum_j\left|\frac{\partial f}{\partial x_j}\right|^2\sigma_{x_j}^2,
\end{equation}
where $\sigma_f$ is the standard deviation of the function $f$, $\sigma_{x_j}$ is the standard deviation of $x_j$. The experimental results of $\mathcal L_\delta^\Delta$ are obtained from the linear function $\mathcal L_\delta^\Delta\equiv\sum_{j=1}^{10}\gamma_j\text{Tr}\left(V_\delta(\vec{\theta})\rho_\text{eq}^\Delta V^\dagger_\delta(\vec{\theta})\sigma_z^j\right)$. In this case,$f=\mathcal L_\delta^\Delta, |\partial f/\partial x_j|^2=\gamma_j^2$ and $x_j=\text{Tr}\left(V_\delta(\vec{\theta})\rho_\text{eq}^\Delta V^\dagger_\delta(\vec{\theta})\sigma_z^j\right)$ is the experimental measurement. The variance of NMR measurement can be calibrated from the SNR in the readout. Due to the polarization of nine identical $^{1}$H spins can be obtained in a single measurement, we have $\sigma^2_{x_1}=0.32,\sum_{j=2}^{10}\sigma^2_{x_j}=0.02$. Hence the standard deviation of experimental $\mathcal L_\delta^\Delta$ is 1.06 and is much smaller than the amplitude of $\mathcal L_\delta^\Delta\sim10^4$.
\end{itemize}
The analysis above is summarized in Table.\ \ref{Tab1}, which shows that the relaxation contributes to $0.98\%$ relative error and is the main source of errors. This result is close to the experimental error of $1.35\%$. 
\begin{table*}
	\caption{Analysis of the experimental errors and the experimental techniques employed to suppress the errors.}\label{Tab1}
	\begin{tabular}{|p{6cm}p{3cm}p{3cm}p{3cm}|}
	\hline
	\multicolumn{4}{|c|}{Totoal experimental error: 1.35$\%$}                                                    \\ \hline
	\multicolumn{1}{|c|}{Sources of experimental error} & \multicolumn{1}{c|}{Pulse error} & \multicolumn{1}{c|}{Relaxation} & \multicolumn{1}{c|}{Measurement error} \\ \hline
	\multicolumn{1}{|c|}{Contributions (regular pulse sequence)} & \multicolumn{1}{c|}{8.1$\%$} & \multicolumn{1}{c|}{8.2$\%$} & \multicolumn{1}{c|}{$2.5\times10^{-2}\%$}\\ \hline
	\multicolumn{1}{|c|}{Contributions} & \multicolumn{1}{c|}{$4.5\times10^{-2}\%$} & \multicolumn{1}{c|}{0.98$\%$} &\multicolumn{1}{c|}{$1.8\times10^{-2}\%$}  \\ 
	\multicolumn{1}{|c|}{(specific experimental technique)} & \multicolumn{1}{c|}{(BB1 sequence)} & \multicolumn{1}{c|}{(calibration with $\mathcal{L}_0^{\Delta}$)} &\multicolumn{1}{c|}{(composite pulse decoupling)}  \\ \hline
	\end{tabular}
\end{table*}

\section{'Time-reversal-based readout' protocol and its precision bound}
The QFI decides the ultimate potential of quantum state, while this bound can only be saturated under optimal measurements. For an arbitrary nonclassical quantum probe, finding optimal measurements is challenging. Inspired by the fact that the 'time-reversal-based readout' can saturate the QCRB for arbitrary pure state \cite{EchoPro}, we generalize this protocol for mixed probe state. Though this protocol doesn't generate optimal measurements, it virtually saturates the QFI under specific parameters. 

The 'time-reversal-based readout' protocol is realized by time reversing the engineering operation $U_\text{E}$ and then project onto the initial state. Consequently, it's similar with the protocol for measuring LE except for replacing the quench $e^{-i\delta G}$ with the realistic encoding process $e^{-i\alpha G}$, where $\alpha$ is the unknown parameter. The precision for estimating $\alpha$ is given by standard error propagation 
\begin{eqnarray}\label{qdelta}
	(\Delta\alpha)^2=\frac{(\Delta\mathcal O)^2}{(d\langle\mathcal O\rangle/d\alpha)^2},
\end{eqnarray}
where $(\Delta\mathcal O)^2:=\langle\mathcal O^2\rangle-\langle\mathcal O\rangle^2$.
For the pure encoded state $e^{i\alpha G}|\Psi_f\rangle$, the QCRB is saturated when $\alpha\to0$ \cite{LEFQ,EchoPro},
\begin{eqnarray}
	(\Delta\alpha)^2=\left.\frac{(\Delta\mathcal O_\text{rev})^2}{(d\langle\mathcal O_\text{rev}\rangle/d\alpha)^2}\right|_{\alpha=0}=\frac{1}{\sqrt{\mathcal F(|\Psi_f\rangle)}}.
\end{eqnarray}
\begin{figure}
	\includegraphics[scale=0.6]{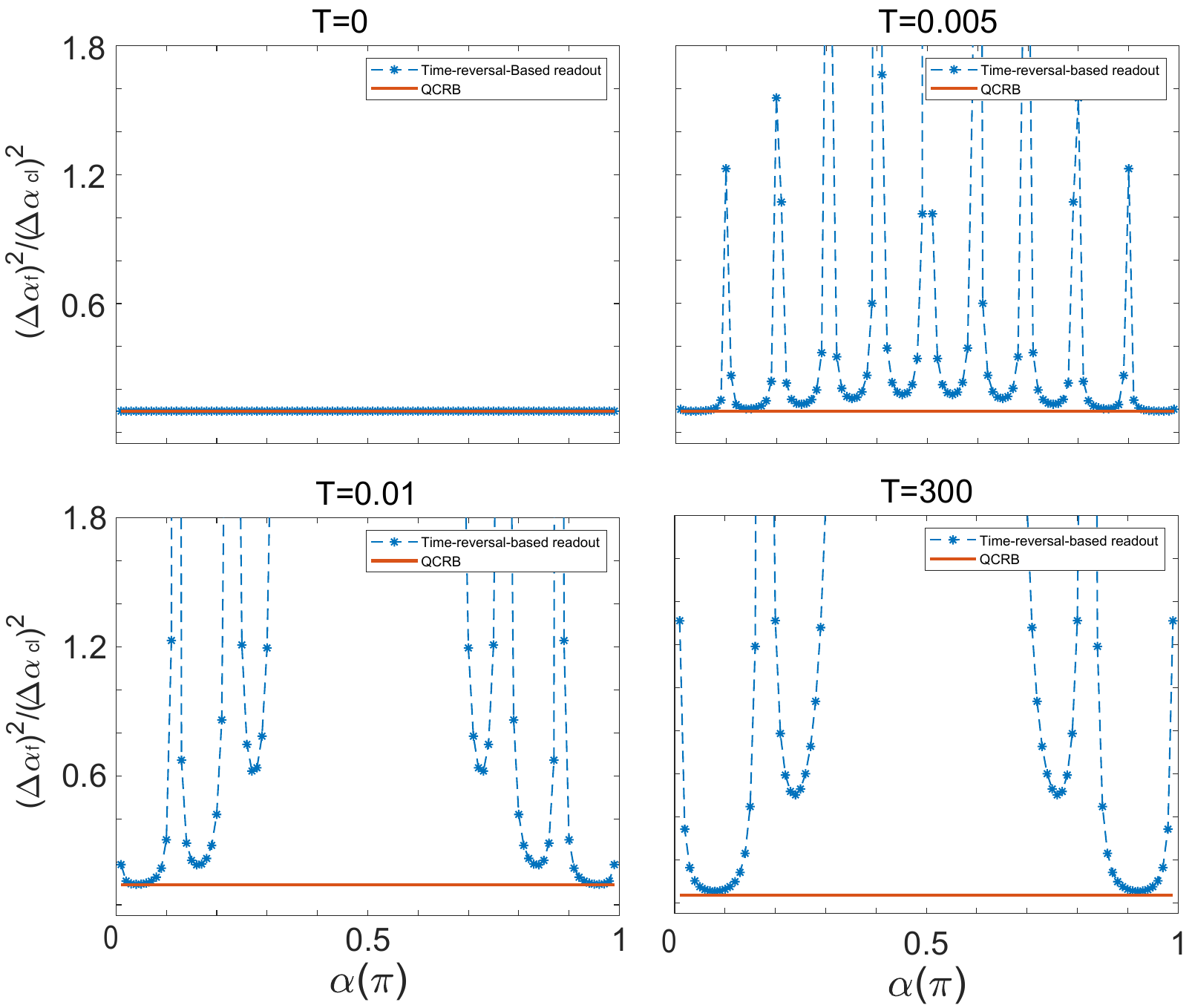} 
	\caption{The ratio of the precision obtained from our 'time-reversal-based readout' protocol to that of standard quantum limit under different purities. Here the purity depends on the Boltzmann distribution at specific temperature. When the temperature is $0$K, the equilibrium state is pure and the precision of 'time-reversal-based readout' protocol can saturate the QCRB. The performance of this protocol becomes worse as the increase of temperature, while a precision close to QCRB is still attainable on some specific region of $\alpha$ even at room temperature.} \label{echo1}
\end{figure}
Here $\mathcal O_\text{rev}=|\Psi_f\rangle\langle\Psi_f|$ and we use the results 
\begin{eqnarray}
	\begin{aligned}
	\Delta\mathcal O_\text{rev}
		=&\langle (|\Psi_f\rangle\langle\Psi_f|)^2\rangle-\langle |\Psi_f\rangle\langle\Psi_f|\rangle^2\\
		=&\frac{\alpha^2}{4}\mathcal F(|\Psi_f\rangle)+O(\alpha^4)\\
	\langle\mathcal O_\text{rev}\rangle=&1-\frac{\alpha^2}{4}\mathcal F(|\Psi_f\rangle)+O(\alpha^4),
	\end{aligned}
\end{eqnarray}
where the expectation is taken over the encoded probe, i.e., $\langle\cdot\rangle:=\langle\Psi_f|e^{i\alpha G}\cdot e^{-i\alpha G}|\Psi_f\rangle$. 

\begin{figure}
	\includegraphics[scale=0.65]{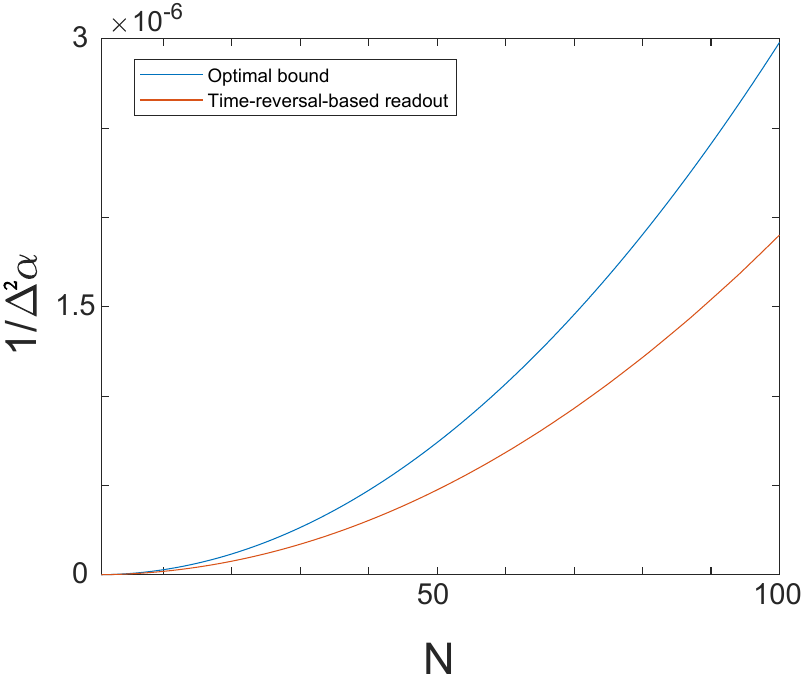} 
	\caption{The scaling of sensitivity under the 'time-reversal-based readout' protocol (red solid line) in terms of the number of particles when $\epsilon=10^{-5}$. It can still achieve the Heisenberg scaling with a ratio to the optimal bound given by Ref. \cite{MixedPRX} (blue solid line) of 0.638.} \label{scaling}
\end{figure}

For the case of mixed probe, the precision $\Delta\alpha$ can also be obtained from Eq. \eqref{qdelta} by calculating $\langle{\mathcal O}_\text{rev}\rangle$ and $\langle{\mathcal O}_\text{rev}^2\rangle$,
\begin{eqnarray}\label{expct}
	\begin{aligned}
		\langle{\mathcal O}_\text{rev}\rangle=&\text{Tr}\left(e^{-i\alpha G}\rho_\text{f} e^{i\alpha G}\rho_\text{f}\right)\\=&\text{Tr}\left(\sum_{m_1}\rho_{m_1}\sum_{m_2}\rho_{m_2}e^{-im_2\alpha}\right)\\
		=&\sum_m\text{Tr}\left(\rho_{-m}\rho_m\right)e^{-im\alpha}\\
		\langle{\mathcal O}^2_\text{rev}\rangle=&\text{Tr}\left(e^{-i\alpha G}\rho_\text{f} e^{i\alpha G}\rho_\text{f}^2\right)\\=&\sum_{m_1,m_2}\text{Tr}\left(\rho_{m_1}\rho_{-m_1-m_2}\rho_{m_2}e^{-im_2\alpha}\right),
	\end{aligned}
\end{eqnarray}
where we divide the density matrix of probe into blocks as $\rho_\text{f}=\sum_m\sum_{\lambda_i-\lambda_j=m}\rho_{ij}|i\rangle\langle j|=\sum_m\rho_m$ with $G|i\rangle=\lambda_i|i\rangle$ \cite{OTOC}. In the following, we specifically consider the initial probe $\rho_0$ in thermal equilibrium, which is close to the case in our experiment. The equilibrium state can be expressed as
\begin{eqnarray}
	\rho_0=(\lambda_0|0\rangle\langle0|+\lambda_1|1\rangle\langle1|)^{\otimes N},
\end{eqnarray}
where $\lambda_0=\frac{e^{\hbar\omega/k_\text{B}T}}{e^{\hbar\omega/k_\text{B}T}+e^{-\hbar\omega/k_\text{B}T}},\lambda_1=\frac{e^{-\hbar\omega/k_\text{B}T}}{e^{\hbar\omega/k_\text{B}T}+e^{-\hbar\omega/k_\text{B}T}}$, $k_\text{B}=1.38\times10^{-23}\text{JK}^{-1}$ is the Boltzmann constant, and $|\hbar\omega|=2.6\times10^{-25}\text{J}$ is the energy difference between the Zeeman states for the case of protons in a field of 9.4T \cite{SpinDyna}. For the optimal probe state engineered by unitary operation from $\rho_0$ (given by Eq. \eqref{optmixed}), the expression of $\langle{\mathcal O}_\text{rev}\rangle$ and $\langle{\mathcal O}_\text{rev}^2\rangle$ become
\begin{eqnarray}
	\begin{aligned}
	\langle{\mathcal O}_\text{rev}\rangle=&\sum_{i=0}^{\left\lfloor\frac{N}{2}\right\rfloor} C_{N}^{i}\left\{\frac{1}{2}\left(\lambda_{0}^{N-i} \lambda_{1}^{i}+\lambda_{0}^{i} \lambda_{1}^{N-i}\right)+\frac{1}{2} \cos [(N-2 i) \alpha]\left(\lambda_{0}^{N-i} \lambda_{1}^{i}-\lambda_{0}^{i} \lambda_{1}^{N-i}\right)\right\},\\
	\langle{\mathcal O}^2_\text{rev}\rangle=
		&\sum_{i=0}^{\left\lfloor\frac{N}{2}\right\rfloor} C_{N}^{i} \frac{1}{2}\left[\left(\lambda_{0}^{N-i} \lambda_{1}^{i}\right)^{2}+\left(\lambda_{0}^{i} \lambda_{1}^{N-i}\right)^{2}\right]\left(\lambda_{0}^{N-i} \lambda_{1}^{i}+\lambda_{0}^{i} \lambda_{1}^{N-i}\right) \\
		&C_{N}^{i} \frac{1}{2} \cos [(N-2 i) \alpha]\left[\left(\lambda_{0}^{N-i} \lambda_{1}^{i}\right)^{2}-\left(\lambda_{0}^{i} \lambda_{1}^{N-i}\right)^{2}\right]\left(\lambda_{0}^{N-i} \lambda_{1}^{i}-\lambda_{0}^{i} \lambda_{1}^{N-i}\right),
		\end{aligned}
\end{eqnarray}
respectively, according to Eq. \eqref{expct}.
Here $C_{N}^{i}$ represents the binomial coefficient. In Fig. \ref{echo1}, we show the precision ratio of optimal engineered state under the 'time-reversal-based readout' protocol to that of classical uncorrelated state in Sec. \ref{secIV} under different temperatures. When the temperature is $0$K, the equilibrium state is pure and the precision under the 'time-reversal-based readout' protocol can saturate the QCRB. The performance of this protocol becomes worse as the increase of temperature, while a precision close to QCRB is still attainable even at room temperature, i.e., the case of our experiments. In this way, we can set the working point as $\widetilde{\alpha}=\text{argmin}_\alpha\Delta\alpha$ for the best metrological performance. The encoded $\alpha$ is priori unknown, while it can be shifted to $\widetilde{\alpha}$ with the adaptive method to saturate the local precision limit \cite{adp1,adp2}.

We further investigate the scaling of the sensitivity of the 'time-reversal-based readout' protocol. Here, the precision bound is obtained at the optimal point $\widetilde{\alpha}$. As shown in Fig. \ref{scaling}, the Heisenberg scaling in terms of the number of particles can still be obtained, whose ratio to the optimal bound in Ref. \cite{MixedPRX} is 0.638.

	\bibliographystyle{apsrev4-2}